\documentclass[pre,twocolumn,showpacs,superscriptaddress,preprintnumbers,floatfix]{revtex4-1}

\usepackage{etex}
\usepackage{ifpdf}
\usepackage{hyperref}
\usepackage{dcolumn}
\usepackage{url}
\usepackage{amsmath}
\usepackage{amscd}
\usepackage{amsfonts}
\usepackage{amssymb}
\usepackage{amsthm}
\usepackage{xfrac}
\usepackage{braket}
\usepackage{bm}   
\usepackage{bbm}
\usepackage{verbatim}
\usepackage{stmaryrd}
\usepackage{xcolor}
\usepackage{setspace}
\usepackage{tikz}
\usepackage[normalem]{ulem}

\usetikzlibrary{arrows}
\usetikzlibrary{automata}
\usetikzlibrary{positioning}
\usetikzlibrary{plotmarks}
\usepackage{pgfplots}
\pgfplotsset{compat=newest}
\usepackage{floatrow}
\tikzstyle{vaucanson}=[
  node distance=2.5cm,
  bend angle=15,
  auto,
  every loop/.style={},
  every edge/.style={->,draw=black,line width=1.2,>=latex,shorten <=1pt,shorten >=1pt},
  every state/.style={draw=black,line width=2,font=\large},
  loop right/.style={right,out=22,in=-22,loop},
  loop above/.style={above,out=112,in=68,loop},
  loop left/.style={left,out=202,in=158,loop},
  loop below/.style={below,out=292,in=248,loop},
  loop above right/.style={right,out=67,in=23,loop},
  loop above left/.style={left,out=157,in=113,loop},
  loop below left/.style={left,out=247,in=203,loop},
  loop below right/.style={right,out=337,in=293,loop},
]

\theoremstyle{plain}    
\theoremstyle{plain}    
\theoremstyle{plain}    
\theoremstyle{plain}    
\theoremstyle{plain}    
\theoremstyle{plain}    
\theoremstyle{plain}    
\theoremstyle{plain}    
\theoremstyle{plain}    
\theoremstyle{plain}    
\theoremstyle{plain}    
\theoremstyle{plain}    
\theoremstyle{plain}    
\theoremstyle{plain}    
\theoremstyle{plain}    
\theoremstyle{plain}    
\theoremstyle{plain}

\newcommand{\eM}     {\mbox{$\epsilon$-machine}}
\newcommand{\eMs}    {\mbox{$\epsilon$-machines}}
\newcommand{\EM}     {\mbox{$\epsilon$-Machine}}
\newcommand{\EMs}    {\mbox{$\epsilon$-Machines}}

\newcommand{\Process}{\mathcal{P}}

\newcommand{\MeasAlphabet}  {\mathcal{A}}
\newcommand{\MeasSymbol}   { {X} }
\newcommand{\meassymbol}   { {x} }
\newcommand{\MeasSymbols}[2]{ \MeasSymbol_{#1:#2} }
\newcommand{\meassymbols}[2] { \meassymbol_{#1:#2} }

\newcommand{\Past} { \MeasSymbols{}{0} }
\newcommand{\past} { \meassymbols{}{0} }

\newcommand{\Future} { \MeasSymbols{0}{} }

\newcommand{\CausalState}   { \mathcal{S} }

\newcommand{\causalstate}   { \sigma }
\newcommand{\CausalStateSet}    { \boldsymbol{\CausalState} }
\newcommand{\AlternateState}    { \mathcal{R} }

\newcommand{\AlternateStateSet} { \boldsymbol{\AlternateState} }

\newcommand{\Prob}      {\Pr} 

\newcommand{\Cmu}       {C_\mu}
\newcommand{\hmu}       {h_\mu}
\newcommand{\EE}        {{\bf E}}

\newcommand{\ProcessAlphabet}   {\MeasAlphabet}

\newcommand{\forward}{+}
\newcommand{\reverse}{-}
\newcommand{\forwardreverse}{\pm} 

\newcommand{\FutureCausalState} { {\CausalState}^{\forward} }

\newcommand{\PastCausalState}   { {\CausalState}^{\reverse} }

\newcommand{\lastindex}[2]{
  \edef\tempa{0}
  \edef\tempb{#2}
  \ifx\tempa\tempb
    \edef\tempc{#1}
  \else
    \edef\tempa{0}
    \edef\tempb{#1}
    \ifx\tempa\tempb
      \edef\tempc{#2}
    \else
      \edef\tempc{#1+#2}
    \fi
  \fi
  \tempc
}

\newcommand{\rmu}{r_\mu}
\newcommand{\bmu}{b_\mu}
\newcommand{\qmu}{q_\mu}

\newcommand{\sigmamu}{\sigma_\mu}

\newcommand{\CSjoint}[1][,]{
   \edef\tempa{:}
   \edef\tempb{#1}
   \ifx\tempa\tempb
      \ensuremath{\FutureCausalState\!#1\PastCausalState}
   \else
      \ensuremath{\FutureCausalState#1\PastCausalState}
   \fi
}

\newif\ifpm
\edef\tempa{\forwardreverse}
\edef\tempb{\pm}
\ifx\tempa\tempb
   \pmfalse
\else
   \pmtrue
\fi

\newcommand{\MIET}{ \langle T\rangle + 1 }

\newcommand{\MISI}{ \frac{1}{\mu} }
\newcommand{\cs} {\causalstate}

\colorlet {R_color}    {blue}
\colorlet {k_color}    {black!30!green}

\def\clap#1{\hbox to 0pt{\hss#1\hss}}

\begin{document}

\title{Time Resolution Dependence of\\
Information Measures for Spiking Neurons:\\
Atoms, Scaling, and Universality
}

\author{Sarah E. Marzen}
\email{smarzen@berkeley.edu}
\affiliation{Department of Physics}

\author{Michael R. DeWeese}
\email{deweese@berkeley.edu}
\affiliation{Department of Physics}
\affiliation{Helen Wills Neuroscience Institute
and Redwood Center for Theoretical Neuroscience\\
University of California at Berkeley, Berkeley, CA 94720}

\author{James P. Crutchfield}
\email{chaos@ucdavis.edu}
\affiliation{Complexity Sciences Center and Department of Physics,\\
University of California at Davis, One Shields Avenue, Davis, CA 95616}

\date{\today}
\bibliographystyle{unsrt}

\begin{abstract}
The mutual information between stimulus and spike-train response is commonly
used to monitor neural coding efficiency, but neuronal computation broadly
conceived requires more refined and targeted information measures of
input-output joint processes. A first step towards that larger goal is to
develop information measures for individual output processes, including
information generation (entropy rate), stored information (statistical
complexity), predictable information (excess entropy), and active information
accumulation (bound information rate). We calculate these for spike trains
generated by a variety of noise-driven integrate-and-fire neurons as a function
of time resolution and for alternating renewal processes. We show that their
time-resolution dependence reveals coarse-grained structural properties of
interspike interval statistics; e.g., $\tau$-entropy rates that diverge less
quickly than the firing rate indicate interspike interval correlations. We also
find evidence that the excess entropy and regularized statistical complexity of
different types of integrate-and-fire neurons are universal in the
continuous-time limit in the sense that they do not depend on mechanism
details. This suggests a surprising simplicity in the spike trains generated by
these model neurons. Interestingly, neurons with gamma-distributed ISIs and
neurons whose spike trains are alternating renewal processes do not fall into
the same universality class. These results lead to two conclusions. First, the
dependence of information measures on time resolution reveals mechanistic
details about spike train generation. Second, information measures can be used
as model selection tools for analyzing spike train processes.

%

\vspace{0.2in}

\end{abstract}

\preprint{Santa Fe Institute Working Paper 15-04-XXX}
\preprint{arxiv.org:1504.XXXX [physics.gen-ph]}

\maketitle


\setstretch{1.1}

\newcommand{\Abet}{\ProcessAlphabet}
\newcommand{\MS}{\MeasSymbol}
\newcommand{\ms}{\meassymbol}
\newcommand{\SSet}{\CausalStateSet}
\newcommand{\St}{\CausalState}
\newcommand{\st}{\causalstate}
\newcommand{\MxSt}{\AlternateState}
\newcommand{\MxSSet}{\AlternateStateSet}
\newcommand{\mxst}{\mu}
\newcommand{\mxstt}[1]{\mu_{#1}}
\newcommand{\StartMS}{\bra{\delta_\pi}}
\newcommand{\Ipred}{\EE}
\newcommand{\ISI} { \phi }


\vspace{0.2in}
\section{Introduction}

Despite a half century of concerted effort~\cite{Mack52a}, neuroscientists
continue to debate the relevant timescales of neuronal communication as well as
the basic coding schemes at work in the cortex, even in early sensory
processing regions of the brain thought to be dominated by feedforward pathways
\cite{SoftkyKoch93,Bell_Mainen_Tsodyks_Sejnowski95,Shadlen_Newsome_95,StevensZador98,Destexhe03,DeWeese_nonGauss06,Koepsell10,Nirenberg09a,Latham10a}.
For example, the apparent variability of neural responses to repeated
presentations of sensory stimuli has led many to conclude that the brain must
average across tens or hundreds of milliseconds or across large populations of
neurons to extract a meaningful signal~\cite{Shadlen_Newsome_98}. Whereas,
reports of reliable responses suggest shorter relevant timescales and more
nuanced coding
schemes~\cite{berry1997structure,Reinagel_2000,DeWeese_binary_03}. In fact,
there is evidence for different characteristic timescales for neural coding in
different primary sensory regions of the cortex~\cite{Yang_Zador_2012}. In
addition to questions about the relevant timescales of neural communication,
there has been an ongoing debate regarding the magnitude and importance of
correlations among the spiking responses of neural
populations~\cite{meister1995concerted,nirenberg2001retinal,schneidman2003synergy,averbeck2006neural,Schneidman_Nature_2006}.

Most studies of neural coding focus on the relationship between a sensory
stimulus and the neural response. Others consider the relationship between the
neural response and the animal's behavioral response \cite{Britten_et_al96},
the relationship between pairs or groups of neurons at different stages of
processing \cite{dan1996efficient,linsker89}, or the variability of neural
responses themselves without regard to other variables
\cite{Schneidman_Nature_2006}. Complementing the latter studies, we are
interested in quantifying the randomness and predictability of neural responses
without reference to stimulus. We consider the variability of a given neuron's
activity at one time and how this is related to the same neuron's activity at
other times in the future and the past.

Along these lines, information theory~\cite{Shan48a,Cove06a} provides an
insightful and rich toolset interpreting neural data and for formulating
theories of communication and computation in the nervous system \cite{Riek99}.
In particular, Shannon's mutual information has developed into a powerful probe
that quantifies the amount of information about a sensory stimulus encoded by
neural
activity~\cite{mac+mc,stein67,atickrev,barlow-redund,laughlin-predcode,laughlin-matchcode,brz,linsker89,sakitt+barlow,Theu91a,frederic,Riek99}.
Similarly, the Shannon entropy has been used to quantify the variability of the
resulting spike-train response. In contrast to these standard stimulus- and
response-averaged quantities, a host of other information-theoretic measures
have been applied in neuroscience, such as the Fisher
information~\cite{Cove06a} and various measures of the information gained per
observation~\cite{DeWeese_Meister,Butts_2006}.

We take an approach that complements more familiar informational analyses.
First, we consider ``output-only'' processes, since their analysis is a
theoretical prerequisite to understanding information in the stimulus-response
paradigm. Second, we analyze rates of informational divergence, not only
nondivergent components. Indeed, we show that divergences, rather than being a
kind of mathematical failure, are important and revealing features of
information processing in spike trains.

We are particularly interested in the information content of neural spiking on
fine timescales. How is information encoded in spike timing and, more
specifically, in interspike intervals? In this regime, the critical questions
turn on determining the kind of information encoded and the required
``accuracy'' of individual spike timing to support it. At present,
unfortunately, characterizing communication at submillisecond time scales and
below remains computationally and theoretically challenging.

Practically, a spike train is converted into a binary sequence for analysis by
choosing a time bin size and counting the number of spikes in successive time
bins. Notwithstanding Refs. \cite{Strong98, Nemenman08a}, there are
few studies of how estimates of communication properties change
as a function of time bin size, though there are examples of both
short~\cite{panzeri1999decoding} and
long~\cite{deweese1996optimization,Strong98} time expansions. Said most
plainly, it is difficult to directly calculate the most basic
quantities---e.g., communication rates between stimulus and spike-train
response---in the submillisecond regime, despite progress on undersampling
\cite{treves1995upward,Nemenman04, Archer12}. Beyond the practical,
the challenges are also conceptual. For example, given that a stochastic process' entropy rate diverges in a process-characteristic fashion for small time discretizations \cite{Gasp93a}, measures of communication efficacy require careful interpretation in this limit.

Compounding the need for better theoretical tools, measurement techniques
will soon amass enough data to allow serious study of neuronal communication at
fine time resolutions and across large populations \cite{Aliv12a}. In this
happy circumstance, we will need guideposts for how information measures of
neuronal communication vary with time resolution so that we can properly
interpret the empirical findings and refine the design of nanoscale probes.

Many single-neuron models generate neural spike trains that are renewal
processes \cite{Gert02a}. Starting from this observation, we use recent results
\cite{Marz14b} to determine how information measures scale in the small
time-resolution limit. This is exactly the regime where numerical methods are
most likely to fail due to undersampling and, thus, where analytic formulae are
most useful. We also extend the previous analyses to structurally more complex,
alternating renewal processes and analyze the time-resolution scaling of their
information measures. This yields important clues as to which scaling results
apply more generally. We then show that, across several standard neuronal
models, the information measures are universal in the sense that their scaling
does not depend on the details of spike-generation mechanisms.

Several information measures we consider are already common fixtures in
theoretical neuroscience, such as Shannon's source entropy rate \cite{Strong98,
Nemenman08a}. Others have appeared at least once, such as the finite-time
excess entropy (or predictive information) \cite{Bial00a,Crut01a} and
statistical complexity \cite{Shaliz10a}. And others have not yet been applied,
such as the bound information \cite{Jame11a,Jame13a}.

The development proceeds as follows. Section \ref{sec:RenewalProcesses} reviews
notation and definitions. To investigate the dependence of causal information
measures on time resolution, Sec. \ref{sec:TimeRes} studies a class of renewal
processes motivated by their wide use in describing neuronal behavior. Section
\ref{sec:GeneralNextSteps} then explores the time-resolution scaling of
information measures of alternating renewal processes, identifying those
scalings likely to hold generally. Section \ref{sec:ComplexityMetrics}
evaluates continuous-time limits of these information measures for common
single-neuron models. This reveals a new kind of universality in which the
information measures' scaling is independent of detailed spiking mechanisms.
Taken altogether, the analyses provide intuition and motivation for several
of the rarely-used, but key informational quantities. For example, the
informational signatures of integrate-and-fire model neurons differ from both
simpler, gamma-distributed processes and more complex, compound renewal
processes. Finally, Sec.  \ref{sec:Conclusions} summarizes the results, giving
a view to future directions and mathematical and empirical challenges.


\section{Background}
\label{sec:RenewalProcesses}

We can only briefly review the relevant physics of information. Much of the
phrasing is taken directly from background presented in Refs. \cite{Marz14a,
Marz14b}.

Let us first recall the causal state definitions \cite{Shal98a} and information
measures of discrete-time, discrete-state processes introduced in Refs.
\cite{Crut08a, Jame11a}. The main object of study is a process $\Process$: the
list of all of a system's behaviors or realizations $\{ \ldots \ms_{-2},
\ms_{-1}, \ms_{0}, \ms_{1}, \ldots \}$ and their probabilities, specified by
the joint distribution $\Prob(\ldots \MS_{-2}, \MS_{-1}, \MS_{0}, \MS_{1},
\ldots)$.  We denote a contiguous chain of random variables as $\MS_{0:L} =
\MS_0 \MS_1 \cdots \MS_{L-1}$. We assume the process is ergodic and
stationary---$\Prob(\MS_{0:L}) = \Prob(\MS_{t:L+t})$ for all $t \in
\mathbb{Z}$---and the measurement symbols range over a finite alphabet: $\ms
\in \Abet$. In this setting, the \emph{present} $\MS_0$ is the random variable
measured at $t = 0$, the \emph{past} is the chain $\MS_{:0} = \ldots \MS_{-2}
\MS_{-1}$ leading up the present, and the \emph{future} is the chain following
the present $\MS_{1:} = \MS_1 \MS_2 \cdots$. (We suppress the infinite index in
these.)

\begin{figure}
\includegraphics[width=\textwidth]{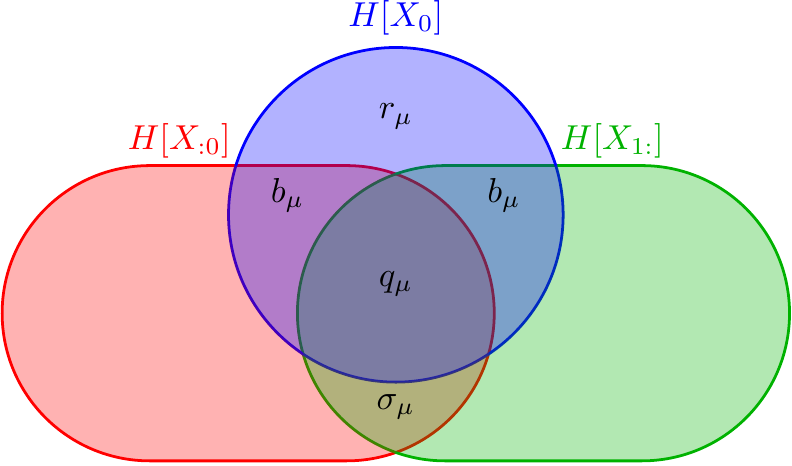}
\caption{Information diagram illustrating the anatomy of the information
  $H[\MS_0]$ in a process' single observation $\MS_0$ in the context of its
  past $\Past$ and its future $\MS_{1:}$. Although the past entropy $H[\Past]$
  and the future entropy $H[\MS_{1:}]$ typically are infinite, space precludes
  depicting them as such. They do scale in a controlled way, however:
  $H[\MS_{-\ell:0}] \propto \hmu \ell$ and $H[\MS_{1:\ell}] \propto \hmu \ell$.
  The two atoms labeled $\bmu$ are the same, since we consider only stationary
  processes. (After Ref. \cite{Jame11a}, with permission.)
  }
\label{fig:IDiagram}
\end{figure}

As the Introduction noted, many information-theoretic studies of neural spike
trains concern input-output information measures that characterize
stimulus-response properties; e.g., the mutual information between stimulus and
resulting spike train. In the absence of stimulus or even with a nontrivial
stimulus, we can still study neural activity from an information-theoretic
point of view using ``output-only'' information measures that quantify
intrinsic properties of neural activity alone:
\begin{itemize}
\setlength{\topsep}{0pt}
\setlength{\itemsep}{0pt}
\setlength{\parsep}{0pt}
\item How random is it? The \emph{entropy rate} $\hmu = H[\MS_0|\MS_{:0}]$,
	which is the entropy in the present observation conditioned on all past
	observations \cite{Cove06a}.
\item What must be remembered about the past to optimally predict the future?
	The \emph{causal states} $\SSet^+$, which are groupings of pasts that lead
	to the same probability distribution over future trajectories
	\cite{Crut88a,Shal98a}.
\item How much memory is required to store the causal states? The
	\emph{statistical complexity} $\Cmu = H[\St^+]$, or the entropy of the
	causal states \cite{Crut88a}.
\item How much of the future is predictable from the past? The
	\emph{excess entropy} $\EE = I[\Past;\MS_{0:}]$, which is the mutual
	information between the past and the future \cite{Crut01a}.
\item How much of the generated information $(h_{\mu})$ is relevant to
	predicting the future? The \textit{bound information} $\bmu =
	I[\MS_0;\MS_{1:}|\MS_{:0}]$, which is the mutual information between the
	present and future observations conditioned on all past observations
	\cite{Jame11a}.
\item How much of the generated information is useless---neither affects future
	behavior nor contains information about the past? The \textit{ephemeral
	information} $\rmu = H[\MS_0|\MS_{:0},\MS_{1:}]$, which is the entropy in
	the present observation conditioned on all past and future observations
	\cite{Jame11a}.
\end{itemize}
The \emph{information diagram} of Fig.~\ref{fig:IDiagram} illustrates the
relationship between $\hmu$, $\rmu$, $\bmu$, and $\EE$. When we change the time
discretization $\Delta t$, our interpretation and definitions change somewhat,
as we describe in Sec.~\ref{sec:TimeRes}.

Shannon's various information quantities---entropy, conditional entropy, mutual
information, and the like---when applied to time series are functions of the
joint distributions $\Prob(\MS_{0:L})$. Importantly, they define an algebra of
information measures for a given set of random variables \cite{Yeun08a}. Ref.
\cite{Jame11a} used this to show that the past and future partition the
single-measurement entropy $H(\MS_0)$ into the measure-theoretic atoms of
Fig.~\ref{fig:IDiagram}. These include those---$\rmu$ and $\bmu$---already
mentioned and the \emph{enigmatic information}:
\begin{align*}
\qmu = I[\MS_0;\MS_{:0};\MS_{1:}] ~,
\end{align*}
which is the co-information between past, present, and future. One can also
consider the amount of predictable information not captured by the present:\begin{align*}
\sigmamu = I[\MS_{:0};\MS_{1:}|\MS_0].
\end{align*}
which is the \emph{elusive information} \cite{Ara14a}. It measures the amount
of past-future correlation not contained in the present. It is nonzero if the
process has ``hidden states'' and is therefore quite sensitive to how the state
space is ``observed'' or coarse-grained.

The total information in the future predictable from the past (or vice
versa)---the excess entropy---decomposes into particular atoms:
\begin{align*}
\EE = \bmu + \sigmamu + \qmu ~.
\end{align*}
The process's Shannon entropy rate $\hmu$ is also a sum of atoms:
\begin{align*}
\hmu = \rmu + \bmu ~.
\end{align*}
This tells us that a portion of the information ($\hmu$) a process spontaneously generates is thrown away ($\rmu$) and a portion is actively stored ($\bmu$). Putting
these observations together gives the information anatomy of a single
measurement $\MS_0$:
\begin{align}
H[\MS_0] = q_{\mu} + 2b_{\mu} + r_{\mu} ~.
\label{eq:SumofRates}
\end{align}
Although these measures were originally defined for stationary processes,
they easily carry over to a nonstationary process of finite Markov order.

Calculating these information measures in closed-form given a model requires
finding the \emph{\eM}, which is constructed from causal states.  Forward-time causal states $\SSet^+$ are minimal sufficient statistics for
predicting a process's future \cite{Crut88a,Shal98a}. This follows from their
definition---a \emph{causal state} $\st^+ \in \SSet^+$ is a sets of pasts grouped by the equivalence relation $\sim^+$:
\begin{align}
\ms_{:0} \sim^+ & \ms_{:0}' \nonumber \\
  & \Leftrightarrow
  \Prob (\MS_{0:}|\MS_{:0}=\ms_{:0}) = \Prob(\MS_{0:}|\MS_{:0}=\ms_{:0}')
  ~.
\end{align}
So, $\SSet^+$ is a set of classes---a coarse-graining of the uncountably
infinite set of all pasts. At time $t$, we have the random variable $\St^+_t$
that takes values $\st^+ \in \SSet^+$ and describes the \emph{causal-state
process} $\ldots, \St^+_{-1}, \St^+_0, \St^+_1, \ldots$. $\St^+_t$ is a
partition of pasts $\MS_{:t}$ that, according to the indexing convention, does
not include the present observation $\MS_t$. In addition to the set of pasts
leading to it, a causal state $\st^+_t$ has an associated \emph{future
morph}---the conditional measure $\Prob(\MS_{t:}| \st^+_t)$ of futures that
can be generated from it. Moreover, each state $\st^+_t$ inherits a probability
$\pi(\st^+_t)$ from the process's measure over pasts $\Prob(\MS_{:t})$. The
forward-time \emph{statistical complexity} is then the Shannon entropy of
the state distribution $\pi(\st^+_t)$ \cite{Crut88a}: $\Cmu^+ = H[\St^+_0]$.
A generative model is constructed out of the causal states by endowing the
causal-state process with transitions:
\begin{align*}
T_{\cs\cs'}^{(\ms)} = \Prob(\St_{t+1}^+=\st',\MS_t = \ms|\St_t^+=\st)
  ~,
\end{align*}
that give the probability of generating the next symbol $\ms$ and ending in
the next state $\cs'$, if starting in state $\cs$. (Residing in a state and
generating a symbol do not occur simultaneously. Since symbols are generated
during transitions there is, in effect, a half time-step difference in the
indexes of the random variables $\MS_t$ and $\St^+_t$. We suppress notating
this.) To summarize, a process's \emph{forward-time \eM} is the tuple
$\{\ProcessAlphabet, \SSet^+, \{ T^{(x)} : \ms \in \ProcessAlphabet \} \}$.

For a discrete-time, discrete-alphabet process, the \eM\ is its minimal
unifilar hidden Markov model (HMM) \cite{Crut88a,Shal98a}. (For general
background on HMMs see \cite{Paz71a,Rabi86a,Rabi89a}.) Note that the causal
state set can be finite, countable, or uncountable; the latter two cases can
occur even for processes generated by finite-state HMMs. \emph{Minimality} can
be defined by either the smallest number of states or the smallest entropy
$H[\St^+_0]$ over states \cite{Shal98a}. \emph{Unifilarity} is a constraint on
the transition matrices $T^{(x)}$ such that the next state $\st'$ is determined
by knowing the current state $\st$ and the next symbol $x$. That is, if the
transition exists, then $\Prob(\St_{t+1}^+|\MS_t = \ms,\St_t^+=\st)$ has
support on a single causal state.

\section{Infinitesimal Time Resolution}
\label{sec:TimeRes}

One often treats a continuous-time renewal process, such as a spike train from
a noisy integrate-and-fire neuron, in a discrete-time setting \cite{Riek99}.
With results of Ref. \cite{Marz14b} in hand, we can investigate how artificial
time binning affects estimates of a model neuron's spike train's randomness,
predictability, and information storage in the limit of infinitesimal time
resolution.  This is exactly the limit in which analytic formulae for
information measures are most useful. For example, as shown shortly in
Fig.~\ref{fig:FofN_IG}, they reveal that increasing the time resolution
artificially increases the apparent range of temporal correlations.

Time-binned neural spike trains of noisy integrate-and-fire neurons have been
studied for quite some time \cite{Mack52a} and, despite that history, this is
still an active endeavor \cite{Riek99,Cess13a}. Our emphasis and approach
differ, though. We do not estimate statistics or reconstruct models from
simulated spike train data using nonparametric inference algorithms---e.g., as
done in Ref. \cite{Shaliz10a}. Rather, we ask how \eMs\ extracted from
a spike train process and information measures calculated from them vary as a
function of time coarse-graining. Our analytic approach highlights an important
lesson about such studies in general: A process' \eM\ and information anatomy
are sensitive to time resolution. A secondary and compensating lesson is that
the manner in which the \eM\ and information anatomy scale with time resolution
conveys much about the process' structure.

Suppose we are given a neural spike train with interspike intervals
independently drawn from the same interspike interval (ISI) distribution
$\ISI(t)$ with mean ISI $1/\mu$. To convert the continuous-time point process
into a sequence of binary spike-quiescence symbols, we track the number of
spikes emitted in successive time bins of size $\Delta t$. Our goal, however,
is to understand how the choice of $\Delta t$ affects reported estimates for
$\Cmu$, $\hmu$, $\EE$, $\bmu$, and $\sigmamu$. The way in which each of these
vary with $\Delta t$ reveals information about the intrinsic time scales on
which a process behaves; cf., the descriptions of entropy rates in Refs.
\cite{costa2002multiscale, costa2005multiscale, Gasp93a}. We concern ourselves
with the infinitesimal $\Delta t$ limit, even though the behavior of these
information atoms is potentially most interesting when $\Delta t$ is on the
order of the process' intrinsic time scales.

In the infinitesimal time-resolution limit, when $\Delta t$ is smaller than any
intrinsic timescale, the neural spike train is a renewal process with
\emph{interevent count distribution}:
\begin{align}
F(n) \approx \ISI(n\Delta t)~\Delta t
\label{eq:ISI_F}
\end{align}
and \emph{survival function}:
\begin{align}
w(n) \approx \int_{n\Delta t}^{\infty} \ISI(t) dt
  ~.
\label{eq:ISI_w}
\end{align}
The interevent distribution $F(n)$ is the probability distribution that the silence separating successive events (bins with spikes) is $n$ counts long. While the survival function $w(n)$ is the probability that the silence separating successive events is at least $n$ counts long.
The \eM\ transition probabilities therefore change
with $\Delta t$. The mean interevent count $\MIET$ is \textit{not} the mean
interspike interval $1/\mu$ since one must convert between counts and spikes
\footnote{As the subscript context makes clear, the mean count $\mu$ is not
related to that $\mu$ in $\Cmu$ and related quantities. In the latter it refers to
the measure over bi-infinite sequences generated by a process.}:
\begin{align}
\MIET = \frac{1}{\mu\Delta t}
  ~.
\label{eq:ISI_T}
\end{align}
In this limit, the \eMs\ of spike-train renewal processes can take one of the
topologies described in Ref. \cite{Marz14b}.

\begin{figure}[htb]
\includegraphics[width=1.05\columnwidth]{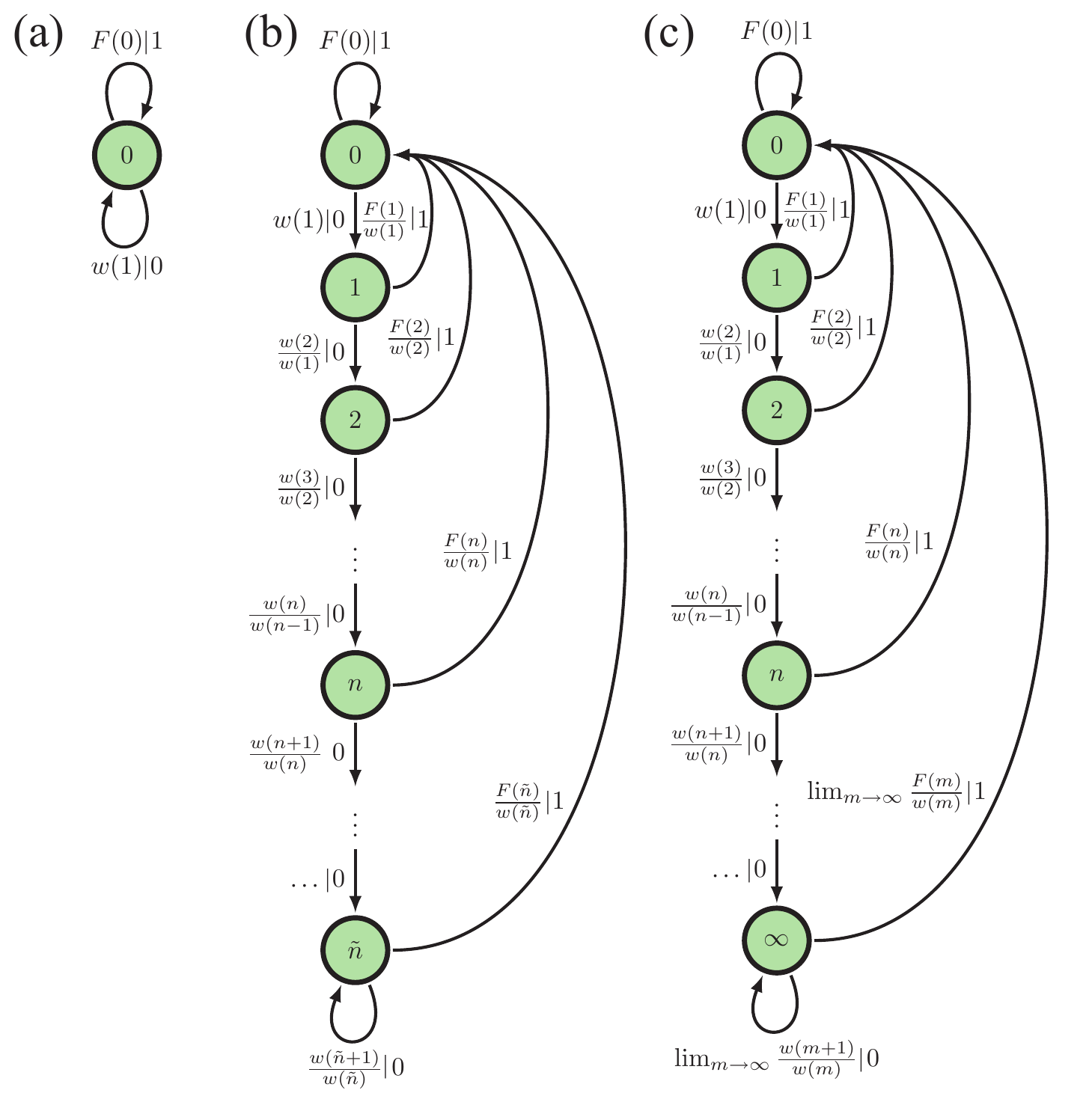}
\caption{\EMs\ of processes generated by Poisson neurons and by
  integrate-and-fire neurons (left to right):
  (a) The \eM\ for a Poisson process.
  (b) The \eM\ for an eventually Poisson process; i.e., a Poisson neuron with a refractory period of length $\tilde{n}\Delta t$.
  (c) The \eM\ for a generic renewal process---the not eventually
  $\Delta$-Poisson process of Ref. \cite{Marz14b}; i.e., the process generated by noise-driven integrate-and-fire neurons.
  Edge labels $p|\ms$ denote emitting symbol $x$ (``1'' is ``spike'') with
  probability $p$.
  (Reprinted with permission from Ref. \protect\cite{Marz14b}.)
  }
\label{fig:Generic_eM}
\end{figure}

\begin{figure}[h!]
\includegraphics[width=\textwidth]{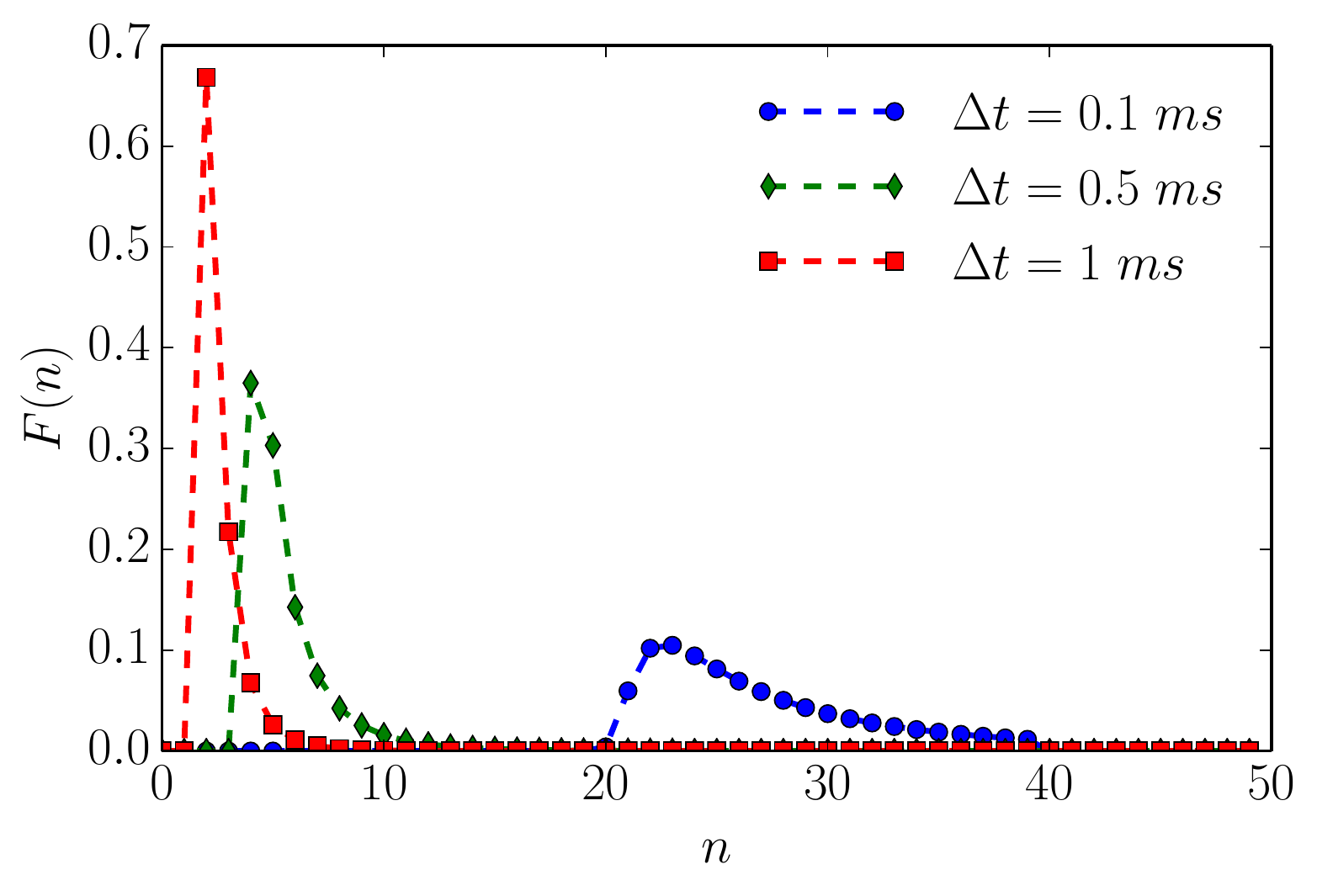}
\caption{An unleaky integrate-and-fire neuron driven by white noise has varying
  interevent count distributions $F(n)$ that depend on time bin size $\Delta t$.
  Based on the ISI distribution $\ISI(t)$ given in Eq.~(\ref{eq:ISI_IG}) with
  $\tau=2$ milliseconds, $1/\mu = 1$ millisecond, and $\lambda=1$ millisecond.
  Data points represent exact values of $F(n)$ calculated for integer values of
  $N$. Dashed lines are interpolations based on straight line segments
  connecting nearest neighbor points.}
\label{fig:FofN_IG}
\end{figure}

Here, we focus only on two of these \eM\ topologies. The first topology
corresponds to that of an eventually Poisson process, in which the ISI
distribution takes the form $\ISI(t) = \ISI(T) e^{-\lambda (t-T)}$ for some
finite $T$ and $\lambda>0$.  A Poisson neuron with firing rate $\lambda$ and
refractory period of time $T$, for instance, eventually ($t > T$) generates an
Poisson process. Hence, we refer to them as \emph{eventually Poisson
processes}. A Poisson process is a special type of eventually Poisson process
with $T=0$; see Fig.~\ref{fig:Generic_eM}(a).  However, the generic renewal
process has \eM\ topology shown in Fig.~\ref{fig:Generic_eM}(c).  Technically,
only noneventually-$\Delta$ Poisson processes have this \eM\ topology, but for
our purposes, this is the \eM\ topology for any renewal process \emph{not}
generated by a Poisson neuron.

At present, inference algorithms can only infer finite \eMs. So, such
algorithms applied to renewal processes will yield an eventually Poisson
topology. (Compare Fig.~\ref{fig:Generic_eM}(c) to the inferred approximate
\eM\ of an integrate-and-fire neuron in Fig.~$2$ in Ref. \cite{Shaliz10a}.) The
generic renewal process has an infinite \eM, though, for which the inferred
\eMs\ are only approximations.

We calculated $\EE$ and $\Cmu$ using the expressions given in Ref.
\cite{Marz14b}.
Substituting in Eqs. (\ref{eq:ISI_F}), (\ref{eq:ISI_w}), and
(\ref{eq:ISI_T}), we find that the excess entropy $\EE$ tends to:
\begin{align}
\lim_{\Delta t\rightarrow 0} \EE(\Delta t)
  & = \int_0^{\infty} \mu t \ISI(t) \log_2
  \big( \mu \ISI(t) \big) dt
  \nonumber \\
  & \quad -2 \int_0^{\infty} \mu \Phi(t)
  \log_2 \big( \mu \Phi(t) \big) dt
  ~,
\label{eq:EE_CT}
\end{align}
where $\Phi(t) = \int_t^{\infty} \ISI(t') dt'$ is the probability that an
ISI is longer than $t$.  It is easy to see that $\EE(\Delta t)$ limits to a positive and (usually) finite value as
the time resolution vanishes, with some exceptions described below.  Similarly, using the expression in Ref.
\cite{Marz14b}'s App. II, one can show that the finite-time excess entropy $\EE
(T)$ \footnote{In the theoretical neuroscience literature, $\EE (T)$ is
sometime called the predictive information $I_{pred}(T)$ and is a useful indicator of process complexity when $\EE$ diverges \cite{Bial00a}.} takes the form:
\begin{align}
\lim_{\Delta t\rightarrow 0}\EE (T)
  & = \left( \int_0^T \mu\Phi(t) dt \right) \log_2 \MISI \nonumber \\
  & \quad\quad\quad - 2\int_0^T \mu \Phi(t)\log_2\Phi(t) dt \nonumber \\
  & \quad\quad\quad - \mu\int_T^{\infty}\Phi(t) dt
     \log_2 \left( \mu \int_T^{\infty}\Phi(t) dt \right) \nonumber \\
  & \quad\quad\quad + \int_0^{T} \mu t F(t)\log_2 F(t) dt \nonumber \\
  & \quad\quad\quad + T \int_T^{\infty} \mu F(t)\log_2 F(t) dt
  ~.
\end{align}
As $T\rightarrow\infty$, $\EE(T)\rightarrow \EE$.  Note that these formulae apply only when mean firing rate $\mu$ is nonzero.

Even if $\EE$ limits to a finite value, the statistical complexity typically
diverges due to its dependence on time discretization $\Delta t$. Suppose that
we observe an eventually Poisson process, such that $\ISI(t)=\ISI(T)
e^{-\lambda (t-T)}$ for $t > T$.
Then, from formulae in Ref. \cite{Marz14b}, statistical complexity in the
infinitesimal time resolution limit becomes:
\begin{align}
\Cmu (\Delta t)
  & \sim \left( \mu \int_0^T \Phi(t) dt \right)
  \log_2 \frac{1}{\Delta t}
  \nonumber \\
   & \quad\quad - \int_0^{T} \left( \mu \Phi(t) \right)
   \log_2 \big( \mu \Phi(t) \big) dt
\label{eq:CmuContTime} \\
   & \quad\quad - \left( \mu \int_T^{\infty} \Phi(t) dt \right)
   \log_2
   \left( \mu \int_T^{\infty} \Phi(t) dt \right) \nonumber
  ~,
\end{align}
ignoring terms of $O(\Delta t)$ or higher. The first term diverges, and its
rate of divergence is the probability of observing a time since last spike less
than $T$. This measures the spike train's deviation from being $\Delta$-Poisson
and so reveals the effective dimension of the underlying causal state space.
$\Cmu$'s remaining nondivergent component is equally interesting.  In fact, it is the
differential entropy of the time since last spike distribution.


An immediate consequence of the analysis is that this generic infinitesimal
renewal process is highly \emph{cryptic} \cite{Crut08a}. It hides an
arbitrarily large amount of its internal state information: $\Cmu$ diverges as
$\Delta t\rightarrow 0$ but $\EE$ (usually) asymptotes to a finite value. We
have very structured processes that have disproportionately little in the
future to predict.
Periodic processes constitute an important exception to this general rule of
thumb for continuous-time processes. A neuron that fires every $T$ seconds without jitter has $\EE=\Cmu$, and both $\EE$ and $\Cmu$ diverge logarithmically with $1/\Delta t$.

It is straightforward to show that any information measure contained within the
present---$H[\MS_0]$, $\hmu$, $\bmu$, $\rmu$, and $\qmu$ (recall Fig.
\ref{fig:IDiagram})---all vanish as $\Delta t$ tends to $0$. Therefore, $\lim_{\Delta t\rightarrow 0} \sigmamu = \lim_{\Delta
t\rightarrow 0} \EE$ and the entropy rate becomes:
\begin{equation}
\hmu \sim -\mu
  \left(\log_2 (\Delta t)
  + \int_0^{\infty} \ISI(t) \log_2 \ISI(t) dt\right) \Delta t
  ~.
\label{eq:hmuContTime}
\end{equation}
With $\Delta t\rightarrow 0$, $\hmu$ nominally tends to $0$: As we shorten the
observation time scale, spike events become increasingly rare. There are at
least two known ways to address $\hmu$ apparently not being very revealing when
so defined. On the one hand, rather than focusing on the uncertainty per
symbol, as $\hmu$ does, we opt to look at the uncertainty per unit time:
$\hmu / \Delta t$. This is the so-called \emph{$\Delta t$-entropy rate}
\cite{Gasp93a} and it diverges as $-\mu \log \Delta t$. Such divergences are
to be expected: The large literature on dimension theory characterizes a
continuous set's
randomness by its divergence scaling rates \cite{Farm83,Maye86a}. Here, we are
characterizing sets of similar cardinality---infinite sequences.
On the other hand, paralleling sequence block-entropy
definition of entropy rate ($\hmu =_{\ell \to \infty}
H[\MS_{0:\ell}] / \ell$) \cite{Crut01a},
continuous-time entropy rates are often approached
within a continuous-time framework using:
\begin{align*}
\hmu = \lim_{T\rightarrow \infty} H(T) / T
  ~,
\end{align*}
where $H(T)$ is path entropy, the continuous-time analogue of the block entropy
$H(\ell)$ \cite{Gira05a}. In these analyses, any $\log\Delta t$ terms are
regularized away using Shannon's differential entropy \cite{Cove06a}, leaving
the nondivergent component $-\mu \int_0^{\infty} \ISI(t)\log \ISI(t) dt$. Using
the $\Delta t$-entropy rate but keeping both the divergent and nondivergent
components, as in Eqs. (\ref{eq:CmuContTime}) and (\ref{eq:hmuContTime}), is an approach that respects both viewpoints and gives a detailed
picture of time-resolution scaling.

A major challenge in analyzing spike trains concerns locating the timescales on
which information relevant to the stimulus is carried.  Or, more precisely, we
are often interested in estimating what percentage of the raw entropy of a
neural spike train is used to communicate information about a stimulus;
cf. the framing in Ref.~\cite{Strong98}. For such analyses, the entropy rate is often taken to be $H(\Delta t,T) / T$, where $T$ is the total path time and $H(\Delta t, T)$ is the entropy of neural spike trains over time $T$ resolved at time bin size $\Delta t$. In terms of previously derived quantities and paralleling the well known block-entropy linear asymptote $H(\ell) = \EE + \hmu \ell$ \cite{Crut01a}, this is:
\begin{align*}
\frac{H(\Delta t,T)}{T}
  = \frac{\hmu(\Delta t)}{\Delta t} + \frac{\EE(T,\Delta t)}{T}
  ~.
\end{align*}
From the scaling analyses above, the extensive component of $H(\Delta t,T)/T$
diverges logarithmically in the small $\Delta t$ limit due to the logarithmic
divergence (Eq. (\ref{eq:hmuContTime})) in $\hmu(\Delta t)/\Delta t$.  If we are interested in accurately estimating the entropy
rate, then the above is one finite-time $T$ estimate of it. However, there
are other estimators, including:
\begin{align*}
\frac{H(\Delta t,T) - H(\Delta t,T-\Delta t)}{\Delta t}
  \approx \frac{\hmu(\Delta t)}{\Delta t}
  + \frac{\partial\EE(T,\Delta t)}{\partial T}
  ~.
\end{align*}
This estimator converges more quickly
to the true entropy rate $\hmu(\Delta t) / \Delta t$ than does $H(\Delta t,T) /
T$.

No such $\log \Delta t$ divergences occur with $\bmu$. Straightforward calculation,
not shown here, reveals that:
\begin{align}
\lim_{\Delta t\rightarrow 0} \frac{\bmu}{\Delta t}
  & = -\mu \left(\int_0^{\infty} \ISI(t)
  \int_0^{\infty} \ISI(t') \log_2 \ISI(t+t')dt' dt
  \right. \nonumber \\
  & \quad\quad\quad \left.
  + \frac{1}{\log 2} - \int_0^{\infty} \ISI(t)\log_2 \ISI(t) dt \right)
  ~. \label{eq:bmu_CT}
\end{align}
Since $\lim_{\Delta t\rightarrow 0} \bmu (\Delta t) / \Delta t < \infty$ and
$\lim_{\Delta t\rightarrow 0} \hmu(\Delta t) / \Delta t$ diverges, the
ephemeral information rate $\rmu (\Delta t) / \Delta t$ also diverges as
$\Delta t\rightarrow 0$. The bulk of the information generated by such renewal
processes is dissipated and, having no impact on future behavior, is not useful
for prediction.

Were we allowed to observe relatively microscopic membrane voltage fluctuations
rather than being restricted to the relatively macroscopic spike sequence, the
$\Delta t$-scaling analysis would be entirely different. Following Ref.
\cite{Marz14a} or natural extensions thereof, the statistical complexity
diverges as $- \log \epsilon$, where $\epsilon$ is the resolution level for the
membrane voltage, the excess entropy diverges as $\log 1 / \Delta t$, the
time-normalized entropy rate diverges as $\log\sqrt{2\pi e D \Delta t} / \Delta
t$, and the time-normalized bound information diverges as $1 / 2\Delta t$. In
other words, observing membrane voltage rather than spikes makes the process
far more predictable. The relatively more macroscopic modeling at the level of
spikes throws away much detail of the underlying biochemical dynamics.

To illustrate the previous points, we turn to
numerics and a particular neural model. Consider an (unleaky) integrate-and-fire neuron driven by white
noise whose membrane voltage (after suitable change of parameters) evolves
according to:
\begin{equation}
\frac{dV}{dt} = b + \sqrt{D}\eta(t) \label{eq:NIF}
  ~,
\end{equation}
where $\eta(t)$ is white noise such that $\langle \eta(t)\rangle = 0$ and
$\langle \eta(t)\eta(t')\rangle = \delta(t-t')$.  When $V=1$, the neuron spikes
and the voltage is reset to $V=0$; it stays at $V=0$ for a time $\tau$, which
enforces a hard refractory period.
Since the membrane voltage resets to a predetermined value, the interspike intervals produced by this model are independently drawn from the same interspike interval distribution:
\begin{equation}
\ISI(t) = \begin{cases}
	0 & t<\tau \\
	\sqrt{\frac{\lambda}{2\pi (t-\tau)^3}}
	e^{-\lambda (\mu (t-\tau)-1)^2/2 (t-\tau)} & t\geq \tau
	\end{cases}
  ~.
\label{eq:ISI_IG}
\end{equation}
Here, $1/\mu = 1 / b$ is the mean interspike interval and $\lambda = 1 / D$ is
a shape parameter that controls ISI variance.
This neural model is not as
realistic as that of a linear leaky integrate-and-fire neural model
\cite{Gert02a}, but is complex enough to illustrate the points made earlier about the scaling of information measures and time resolution.

\begin{figure*}[tb]
\includegraphics[width=\textwidth]{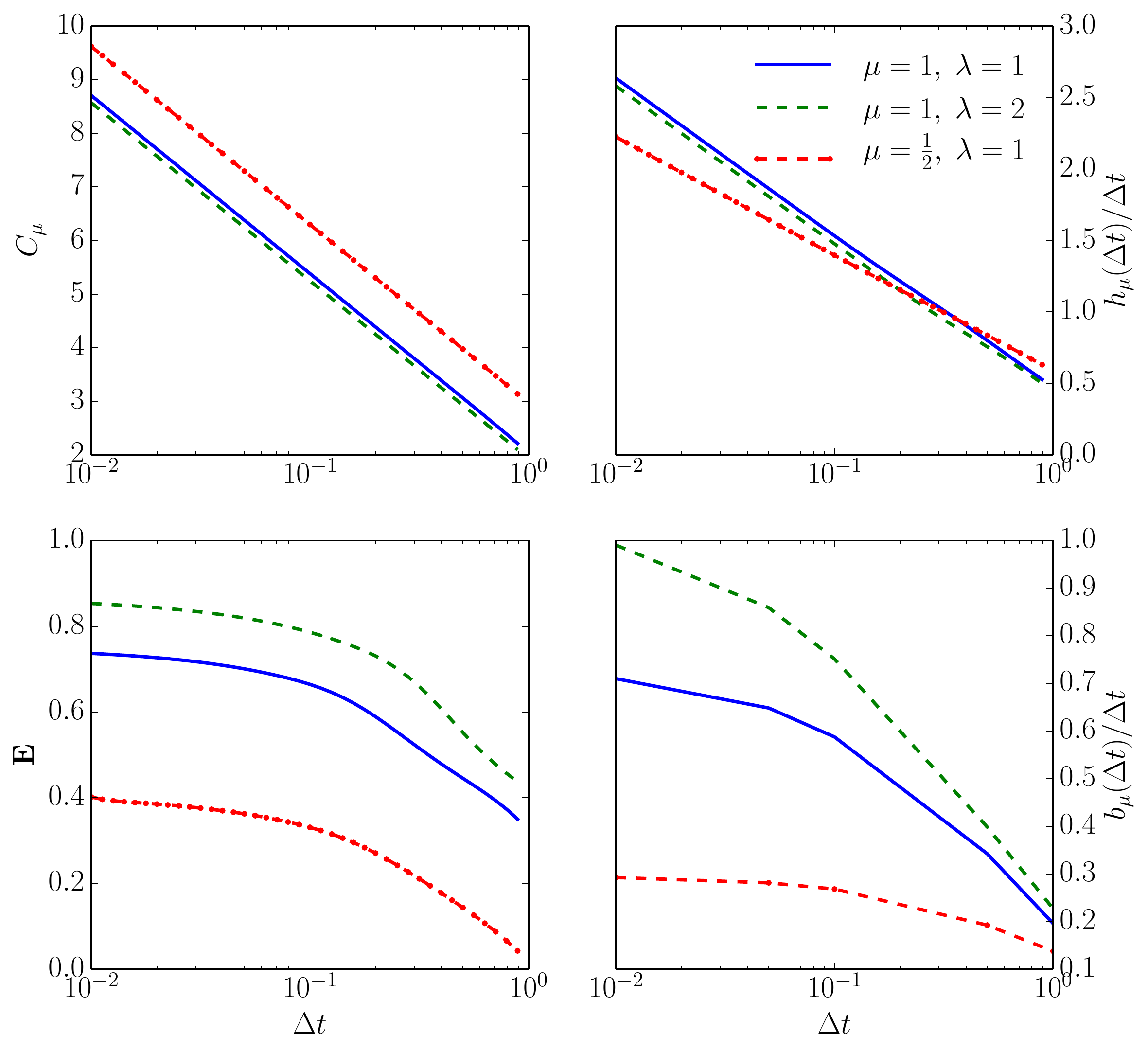}
\caption{How spike-train information measures (or rates) depend on time
  discretization $\Delta t$ for an unleaky integrate-and-fire neuron driven
  by white noise.
Top left: Statistical complexity $\Cmu$ as a function of both the ISI
  distribution shape parameters and the time bin size $\Delta t$. The
  horizontal axis is $\Delta t$ in milliseconds on a log-scale and the vertical
  axis is $\Cmu$ in bits on a linear scale for three different ISI distributions
  following Eq.~(\ref{eq:ISI_IG}) with $\tau =2$ milliseconds.
 Top right:
  Entropy rate $\hmu$ also as a function of both shape parameters and $\Delta
  t$. Axes labeled as in the previous panel and the same three ISI
  distributions are used. Bottom left: Excess entropy $\EE$ as a function of
  both the shape parameters and $\Delta t$. For the blue line $\lim_{\Delta
  t\rightarrow 0}\EE(\Delta t) = 0.75$ bits; purple line, $\lim_{\Delta
  t\rightarrow 0} \EE(\Delta t) = 0.86$ bits; and yellow line, $\lim_{\Delta
  t\rightarrow 0} \EE(\Delta t) = 0.41$ bits. All computed from
  Eq.~(\ref{eq:EE_CT}). Bottom right: Bound information rate $\bmu(\Delta
  t)/\Delta t$ parametrized as in the previous panels. For the blue line
  $\lim_{\Delta t\rightarrow 0} \bmu(\Delta t)/\Delta t = 0.73$ bits per
  second; purple line, $\lim_{\Delta t\rightarrow 0} \bmu(\Delta t)/\Delta t =
  1.04$ bits per second; and yellow line, $\lim_{\Delta t\rightarrow 0}
  \bmu(\Delta t)/\Delta t = 0.30$ bits per second.  All computed from
  Eq.~\ref{eq:bmu_CT}.
  }
\label{fig:FiniteTimeRes}
\end{figure*}

For illustration purposes, we assume that the time-binned neural spike train is
well approximated by a renewal process, even when $\Delta t$ is as large as one
millisecond. This assumption will generally not hold, as past interevent counts
could provide more detailed historical information that more precisely places
the last spike within its time bin. Even so, the reported information measure
estimates are still useful. The estimated $\hmu$ is an upper bound on the true
entropy rate; the reported $\EE$ is a lower bound on the true excess entropy
using the Data Processing Inequality \cite{Cove06a}; and the reported $\Cmu$
will usually be a lower bound on the true process' statistical complexity.

Employing the renewal process assumption, numerical analysis corroborates the
infinitesimal analysis above. Figure \ref{fig:FofN_IG} plots $F(n)$---the proxy
for the full, continuous-time, ISI distribution---for a given set of neuronal
parameter values as a function of time resolution. Figure
\ref{fig:FiniteTimeRes} then shows that $\hmu$ and $\Cmu$ exhibit
logarithmic scaling at millisecond time discretizations, but that $\EE$ does
not converge to its continuous-time value until we reach time discretizations
on the order of hundreds of microseconds. Even when $\Delta
t=100~\mu s$, $\bmu(\Delta t)/\Delta t$ still has not converged to its
continuous-time values.

The statistical complexity $\Cmu$ increases without bound, as $\Delta t \to 0$;
see the top left panel of Fig. \ref{fig:FiniteTimeRes}. As suggested in the
infinitesimal renewal analysis, $\hmu$ vanishes, whereas $\hmu / \Delta t$
diverges at a rate of $\mu \log_2 1/\Delta t$, as shown in the top right plots
of Fig. \ref{fig:FiniteTimeRes}. As anticipated, $\EE$ tends to a finite, ISI
distribution-dependent value when $\Delta t$ tends to $0$, as shown in the
bottom left panel in Fig. \ref{fig:FiniteTimeRes}. Finally, the lower right
panel plots $\bmu(\Delta t)/\Delta t$.

One conclusion from this simple numerical analysis is that one should consider
going submillisecond time resolutions to obtain accurate estimates of
$\lim_{\Delta t\rightarrow 0}\EE(\Delta t)$ and $\lim_{\Delta t\rightarrow 0}
\bmu(\Delta t)/\Delta t$, even though the calculated informational values are a
few bits or even less than one bit per second in magnitude.

\begin{figure*}[htb]
\includegraphics[width=\textwidth]{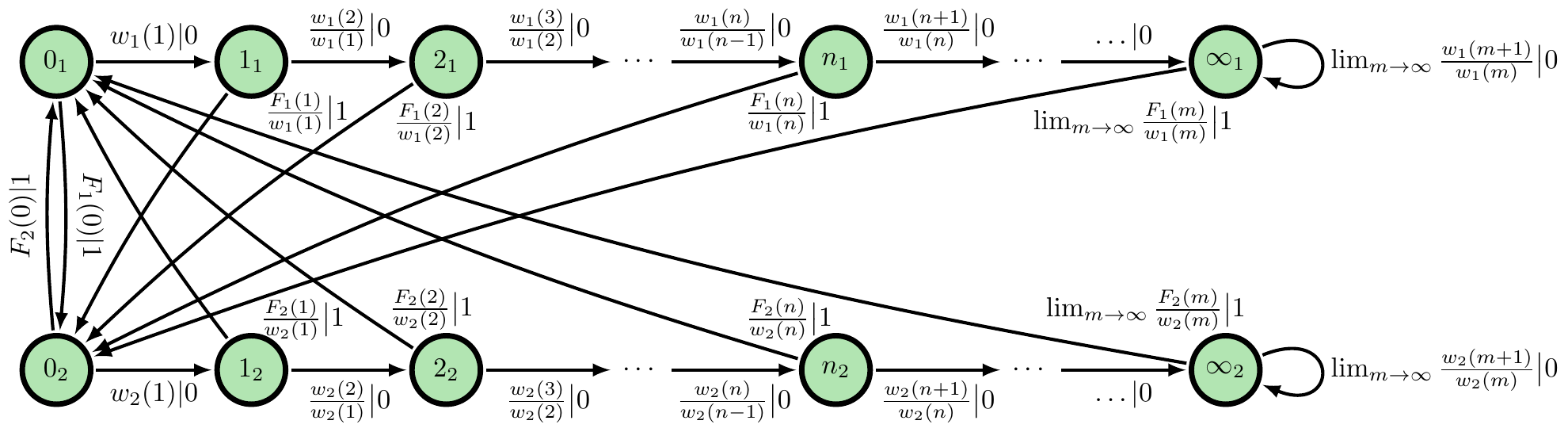}
\caption{\EM\ for an alternating renewal process in which neither interevent
  count distribution is $\Delta$-Poisson and they are not equal almost
  everywhere. State label $n_m$ denotes $n$ counts since the last event and
  present modality $m$.
  }
\label{fig:ARPeM}
\end{figure*}

\section{Alternating renewal processes}
\label{sec:GeneralNextSteps}

The form of the $\Delta t$-scalings discussed in Sec. \ref{sec:TimeRes} occur
much more generally than indicated there. Often, our aim is to calculate the
nondivergent component of these information measures as $\Delta t\rightarrow
0$, but the rates of these scalings are process-dependent. Therefore, these
divergences can be viewed as a feature rather than a bug; they contain
additional information about the process' structure \cite{Gasp93a}.

To illustrate this point, we now investigate $\Delta t$-scalings for
information measures of alternating renewal processes (ARPs), which are
structurally more complex than the standard renewal processes considered above.
For instance, these calculations suggest that rates of divergence of the
$\tau$-entropy rate smaller than the firing rate, such as those seen in Ref.
\cite{Nemenman08a}, are indicative of strong ISI correlations.  Calculational
details are sequestered in App. \ref{app:ARPs}.

In an ARP, an ISI is drawn from one distribution $\ISI^{(1)}(t)$, then another
distribution $\ISI^{(2)}(t)$, then the first $\ISI^{(1)}(t)$ again, and so on.
We refer to the new piece of additional information---the ISI distribution
currently being drawn from---as the \emph{modality}. Under weak technical
conditions, the causal states are the modality and time since last spike.  The
corresponding, generic \eM\ is shown in Fig. \ref{fig:ARPeM}. We define the
modality-dependent survival functions as $\Phi_i(t) =
\int_t^{\infty}\ISI^{(i)}(t') dt'$, the modality-dependent mean firing rates as:
\begin{align}
\mu^{(i)} = 1 \Big/ \int_0^{\infty} \ISI^{(i)} (t) dt
  ~,
\label{eq:ARPmu}
\end{align}
the modality-dependent differential entropy rates:
\begin{align*}
\hmu^{(i)} = -\mu^{(i)} \int_0^{\infty} \ISI^{(i)}\log_2\ISI^{(i)}(t) dt
  ~,
\end{align*}
the modality-dependent continuous-time statistical complexity:
\begin{align*}
\Cmu^{(i)} = -\int_0^{\infty} \mu^{(i)} \Phi^{(i)}(t)\log_2
  \left( \mu^{(i)}\Phi^{(i)}(t) \right) dt
  ~,
\end{align*}
and the modality-dependent excess entropy:
\begin{align}
\EE^{(i)} & = \int_0^{\infty} \mu^{(i)} t \ISI^{(i)}(t)
  \log_2 \left( \mu^{(i)}\ISI^{(i)}(t) \right) dt \nonumber \\
  & \quad\quad - 2\int_0^{\infty} \mu^{(i)}
  \Phi^{(i)}(t) \log_2 \left( \mu^{(i)}\Phi^{(i)}(t) \right) dt
  ~.
\label{eq:ARPEE}
\end{align}

It is straightforward to show, as done in App. \ref{app:ARPs}, that the
time-normalized entropy rate still scales with $\log_2 1 / \Delta t$:
\begin{align}
\frac{\hmu(\Delta t)}{\Delta t}
  \!\sim\! \frac{\mu^{(1)}\mu^{(2)}}{\mu^{(1)}+\mu^{(2)}}
  \log_2\left(\frac{1}{\Delta t}\right)
  \!+\!
  \frac{\mu^{(2)}\hmu^{(1)} \!+\! \mu^{(1)}\hmu^{(2)}}{\mu^{(1)}+\mu^{(2)}}
  .
\label{eq:ARP1}
\end{align}
As expected, the statistical complexity still diverges:
\begin{align}
\Cmu(\Delta t) & \sim 2\log_2\left(\frac{1}{\Delta t}\right)
  + \frac{\mu^{(2)}\Cmu^{(1)}+\mu^{(1)}\Cmu^{(2)}}{\mu^{(1)}+\mu^{(2)}}
  \nonumber \\
  & \quad\quad + H_b \left( \frac{\mu_1}{\mu_1+\mu_2} \right)
  ~,
\label{eq:ARPCmu}
\end{align}
where $H_b(p) = -p\log_2 p - (1-p)\log_2 (1-p)$ is the entropy in bits of a
Bernoulli random variable with bias $p$. Finally, the excess entropy still limits to a positive constant:
\begin{align}
\lim_{\Delta t\rightarrow 0} \EE(\Delta t)
  = H_b \left( \frac{\mu_1}{\mu_1+\mu_2} \right)
  + \frac{\mu^{(2)}\EE^{(1)}+\mu^{(1)}\EE^{(2)}}{\mu^{(1)}+\mu^{(2)}}
  ~.
  \label{eq:ARPEnd}
\end{align}
The additional terms $H_b(\cdot)$ come from the information stored in the time course of modalities.

As a point of comparison, we ask what these information measures would be for
the original (noncomposite) renewal process with the same ISI distribution as
the ARP. As described in App. \ref{app:SiCM}, the former entropy rate is
always less than the true $\hmu$; its statistical complexity is always less
than the true $\Cmu$; and its excess entropy is always smaller than the true
$\EE$. In particular, the ARP's $\hmu$ divergence rate is always less
than or equal to the mean firing rate $\mu$. Interestingly, this coincides with
what was found empirically in the time series of a single neuron; see Fig. 5C
in Ref.  \cite{Nemenman08a}.

The ARPs here are a first example of how one
can calculate information measures of the much
broader and more structurally complex class of processes generated by unifilar
hidden semi-Markov models, a subclass of hidden semi-Markov models
\cite{Tokdar10}.

\section{Information Universality}
\label{sec:ComplexityMetrics}

Another aim of ours was is to interpret the information measures. In
particular, we wished to relate infinitesimal time-resolution excess entropies,
statistical complexities, entropy rates, and bound information rates to more
familiar characterizations of neural spike trains---firing rates $\mu$ and ISI
coefficient of variations $C_V$. To address this, we now analyze a suite of
familiar single-neuron models. We introduce the models first, describe the
parameters behind our numerical estimates, and then compare the information
measures.

Many single-neuron models, when driven by temporally uncorrelated and
stationary input, produce neural spike trains that are renewal processes. We
just analyzed one model class, the noisy integrate-and-fire (NIF) neurons in Sec.
\ref{sec:TimeRes}, focusing on time-resolution dependence. Other common
neural models include the linear leaky integrate-and-fire (LIF) neuron, whose
dimensionless membrane voltage, after a suitable change of parameters,
fluctuates as:
\begin{align}
\frac{dV}{dt} = b -V + a\eta(t)
  ~,
\label{eq:NLLIF}
\end{align}
and when $V=1$, a spike is emitted and $V$ is instantaneously reset to $0$. We
computed ISI survival functions from empirical histograms of $10^5$ ISIs; we
varied $b \in [1.5,5.75]$ in steps of $0.25$ and $a \in [0.1,3.0]$ in steps of
$0.1$ to $a = 1.0$ and in steps of $0.25$ thereafter.

The quadratic integrate-and-fire (QIF) neuron has membrane voltage fluctuations
that, after a suitable change of variables, are described by:
\begin{align}
\frac{dV}{dt} = b + V^2 + a\eta(t)
  ~,
\label{eq:NQIF}
\end{align}
and when $V=100$, a spike is emitted and $V$ is instantaneously reset to
$-100$. We computed ISI survival functions from empirical histograms of
trajectories with $10^5$ ISIs; we varied $b \in [0.25,4.75]$ in steps of $0.25$
and $a \in [0.25,2.75]$ in steps of $0.25$. The QIF neuron has a very different
dynamical behavior from the LIF neuron, exhibiting a Hopf bifurcation at $b=0$.
Simulation details are given in App. \ref{app:SiCM}.

Finally, ISI distributions are often fit to gamma distributions, and so we also
calculated the information measures of spike trains with gamma-distributed ISIs
(GISI).

\begin{figure*}
\includegraphics[width=\textwidth]{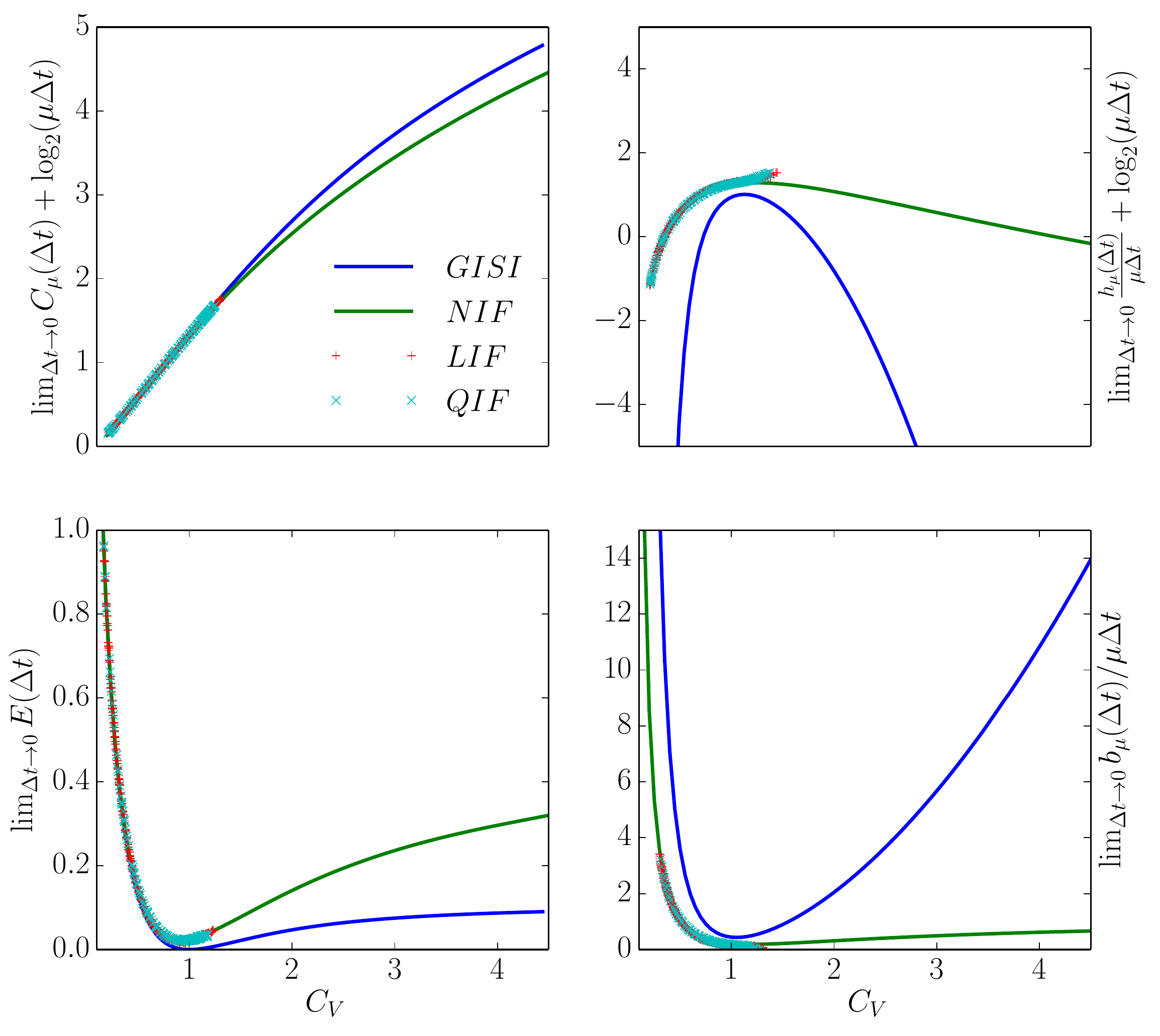}
\caption{Information universality across distinct neuron dynamics. We find that
  several information measures depend only on the ISI coefficient of variation
  $C_V$ and not the ISI mean firing rate $\mu$ for the following neural spike
  train models:
  (i) neurons with Gamma distributed ISIs (GISI, blue), (ii) noisy
  integrate-and-fire neurons governed by Eq.~(\ref{eq:NIF}) (NIF, green), (iii)
  noisy linear leaky integrate-and-fire neurons governed by
  Eq.~(\ref{eq:NLLIF}) (LIF, dotted red), and (iv) noisy quadratic
  integrate-and-fire neurons governed by Eq.~(\ref{eq:NQIF}) (QIF, dotted
  blue). Top left: $\lim_{\Delta t\rightarrow 0}\Cmu(\Delta t)+\log_2(\mu\Delta
  t)$. Top right: $\lim_{\Delta t\rightarrow 0} \hmu(\Delta t)/\mu\Delta
  t+\log_2(\mu\Delta t)$. Bottom left: $\lim_{\Delta t\rightarrow 0}\EE(\Delta
  t)$. Bottom right: $\lim_{\Delta t\rightarrow 0} \bmu(\Delta t)/\mu\Delta
  t)$. In the latter, ISI distributions with smaller $C_V$ were excluded due to
  the difficulty of accurately estimating $\int_0^{\infty} \int_0^{\infty}
  \ISI(t)\ISI(t') \log_2 \ISI(t+t') dt dt'$ from simulated spike trains. See
  text for discussion.
}
\label{fig:Universality}
\end{figure*}

Each neural model---NIF, LIF, QIF, and GISI---has its own set of parameters
that governs its ISI distribution shape. Taken at face value, this would make
it difficult to compare information measures across models. Fortunately, for
each of these neural models, the firing rate $\mu$ and coefficient of variation
$C_V$ uniquely determine the underlying model parameters \cite{Lindner09a}.
As App. \ref{app:SiCM} shows, the quantities $\lim_{\Delta
t\rightarrow 0}\EE(\Delta t)$, $\lim_{\Delta t\rightarrow 0} \Cmu+\log_2
(\mu\Delta t)$, $\lim_{\Delta t\rightarrow 0} \hmu(\Delta t)/\mu\Delta t +
\log_2 (\mu\Delta t)$, and $\lim_{\Delta t\rightarrow 0} \bmu(\Delta
t)/\mu\Delta t$ depend only on the ISI coefficient of variation $C_V$ and not
the mean firing rate $\mu$.

We estimated information measures from the simulated spike train data using
plug-in estimators based on the formulae in Sec. \ref{sec:TimeRes}. Enough data
was generated that even naive plug-in estimators were adequate \emph{except}
for estimating $\bmu$ when $C_V$ was larger than $1$. See App. \ref{app:SiCM}
for estimation details. That said, binned estimators are likely inferior to
binless entropy estimators \cite{Victor02a}, and naive estimators tend to have
large biases. This will be an interesting direction for future research, since
a detailed analysis goes beyond the present scope.

Figure \ref{fig:Universality} compares the statistical complexity, excess
entropy, entropy rate, and bound information rate for all four neuron types as
a function of their $C_V$. Surprisingly, the NIF, LIF, and QIF neuron's
information measures have essentially identical dependence on $C_V$. That is,
the differences in mechanism do not strongly affect these informational
properties of the spike trains they generate. Naturally, this leads one to ask
if the informational indifference to mechanism generalizes to other spike train
model classes and stimulus-response settings.

Figure \ref{fig:Universality}'s top left panel shows that the continuous-time
statistical complexity grows monotonically with increasing $C_V$. In
particular, the statistical complexity increases logarithmically with ISI mean
and approximately linearly with the ISI coefficient of variation $C_V$. That
is, the number of bits that must be stored to predict these processes increases
in response to additional process stochasticity and longer temporal
correlations. In fact, it is straightforward to show that the statistical
complexity is minimized and excess entropy maximized at fixed $\mu$ when the
neural spike train is periodic. This is unsurprising since, in the space of
processes, periodic processes are least cryptic ($\Cmu - \EE = 0$) and so
knowledge of oscillation phase is enough to completely predict the future. (See
App. \ref{app:SiCM}.)

The bottom left panel in Figure \ref{fig:Universality} shows that increasing
$C_V$ tends to decrease the excess entropy $\EE$---the number of bits that one
can predict about the future. $\EE$ diverges for small $C_V$, dips at the $C_V$
where the ISI distribution is closest to exponential, and limits to a small
number of bits at large $C_V$. At small $C_V$, the neural spike train is close
to noise-free periodic behavior. When analyzed at small but nonzero $\Delta t$,
$\EE$ encounters an ``ultraviolet divergence'' \cite{Tche13a}. Thus, $\EE$
diverges as $C_V\rightarrow 0$, and a simple argument in App. \ref{app:SiCM}
suggests that the rate of divergence is $\log_2 (1 / C_V)$. At an intermediate
$C_V \sim 1$, the ISI distribution is as close as possible to that of a
memoryless Poisson process and so $\EE$ is close to vanishing. At larger $C_V$,
the neural spike train is noise-driven. Surprisingly, completely noise-driven
processes still have a fraction of a bit of predictability: knowing the time since last spike allows for some power in predicting the time to next spike.

The top right panel shows that an appropriately rescaled differential entropy rate varies differently for neural spike trains from noisy integrate-and-fire neurons and neural spike trains with gamma-distributed ISIs.  As expected, the entropy rate is maximized at $C_V$ near $1$, consistent with the Poisson process being the maximum entropy distribution for fixed mean ISI.  Gamma-distributed ISIs are far less random than ISIs from noisy integrate-and-fire neurons, holding $\mu$ and $C_V$ constant.

Finally, the continuous-time bound information ($\bmu$) rate varies in a
similar way to $\EE$ with $C_V$. (Note that since the plotted quantity is
$\lim_{\Delta t\rightarrow 0} \bmu(\Delta t)/\mu\Delta t$, one could interpret
the normalization by $1/\mu$ as a statement about how the mean firing rate
$\mu$ sets the natural timescale.) At low $C_V$, the $\bmu$ rate diverges as
$1/C_V^2$, as described in App. \ref{app:SiCM}. Interestingly, this limit is
singular, similar to the results in Ref. \cite{Marz14a}: at $C_V=0$, the spike
train is noise-free periodic and so the $\bmu$ rate is $0$. For $C_V\approx 1$,
it dips for the same reason that $\EE$ decreases. For larger $C_V$, $\bmu$'s
behavior depends rather strongly on the ISI distribution shape. The
longer-ranged gamma-distribution results in ever-increasing $\bmu$ rate for
larger $C_V$, while the $\bmu$ rate of neural spike trains produced by NIF
neurons tends to a small positive constant at large $C_V$. The variation of
$\bmu$ deviates from that of $\EE$ qualitatively at larger $C_V$ in that the
GISI spike trains yield smaller total predictability $\EE$ than that of NIF
neurons, but arbitrarily higher predictability \emph{rate}.

These calculations suggest a new kind of universality for neuronal information
measures \emph{within a particular generative model class}. All of these
distinct integrate-and-fire neuron models generate ISI distributions from
different families, yet their informational properties exhibit the same
dependencies on $\Delta t$, $\mu$, and $C_V$ in the limit of small $\Delta t$.
Neural spike trains with gamma-distributed ISIs did not show similar
informational properties. And, we would not expect neural spike trains that are
alternating renewal processes to show similar informational properties either.
(See Sec.~\ref{sec:GeneralNextSteps}.)  These coarse information quantities
might therefore be effective model selection tools for real neural spike train
data, though more groundwork must be explored to ascertain their utility.


\section{Conclusions}
\label{sec:Conclusions}

We explored the scaling properties of a variety of information-theoretic
quantities associated with two classes of spiking neural models: renewal
processes and alternating renewal processes. We found that information
generation (entropy rate) and stored information (statistical complexity) both
diverge logarithmically with decreasing time resolution for both types of
spiking models, whereas the predictable information (excess entropy) and active
information accumulation (bound information rate) limit to a constant. Our
results suggest that the excess entropy and regularized statistical complexity
of different types of integrate-and-fire neurons are universal in the sense
that they do not depend on mechanism details, indicating a surprising
simplicity in complex neural spike trains. Our findings highlight the
importance of analyzing the scaling behavior of information quantities, rather
than assessing these only at a fixed temporal resolution.

By restricting ourselves to relatively simple spiking models we have been able
to establish several key properties of their behavior. There are, of
course, other important spiking models that cannot be expressed as renewal
processes or alternating renewal processes, but we are encouraged by the robust
scaling behavior of the entropy rate, statistical complexity, excess entropy,
and bound information rate over the range of models we considered.

There was a certain emphasis here on the entropy rate and hidden Markov models
of neural spike trains, both familiar tools in computational neuroscience. On
this score, our contributions are straightforward. We determined how the
entropy rate varies with the time discretization and identified the possibly
infinite-state, unifilar HMMs required for optimal prediction of spike-train
renewal processes.  Entropy rate diverges logarithmically for stochastic
processes \cite{Gasp93a}, and this has been observed empirically for neural
spike trains for time discretizations in the submillisecond regime
\cite{Nemenman08a}. We argued that the $\hmu$ divergence rate is an important
characteristic. For renewal processes, it is the mean firing rate; for
alternating renewal processes, the ``reduced mass'' of the mean firing rates.
Our analysis of the latter, more structured processes showed that a divergence
rate less than the mean firing rate---also seen experimentally
\cite{Nemenman08a}---indicates that there are strong correlations between ISIs.
Generally, the nondivergent component of the time discretization-normalized
entropy rate is the differential entropy rate; e.g., as given in Ref.
\cite{Stev96a}.

Empirically studying information measures as a function of time resolution can
lead to a refined understanding of the time scales over which neuronal
communication occurs. Regardless of the information measure chosen, the results
and analysis here suggest that much can be learned by studying scaling behavior
rather than focusing only on neural information as a single quantity estimated
at a fixed temporal resolution. While we focused on the regime in which the
time discretization was smaller than any intrinsic timescale of the process,
future and more revealing analyses would study scaling behavior at even smaller
time resolutions to directly determine intrinsic time scales \cite{Crut92c}.

Going beyond information generation (entropy rate), we analyzed information
measures---namely, statistical complexity and excess entropy---that have only
recently been used to understand neural coding and communication.  Their
introduction is motivated by the hypothesis that neurons benefit from learning
to predict their inputs \cite{Palm13a}, which can consist of the neural spike
trains of upstream neurons. The statistical complexity is the minimal amount of
historical information required for exact prediction. To our knowledge, the
statistical complexity has appeared only once previously in computational
neuroscience \cite{Shaliz10a}. The excess entropy, a closely related companion,
is the maximum amount of information that can be predicted about the future.
When it diverges, then its divergence rate is quite revealing of the underlying
process \cite{Crut92c,Bial00a}, but none of the model neural spike trains
studied here had divergent excess entropy. Finally, the bound information rate
has yet to be deployed in the context of neural coding, though related
quantities have drawn attention elsewhere, such as in nonlinear dynamics
\cite{Jame13a} and information-based reinforcement learning \cite{Mart13a}.
Though its potential uses have yet to be exploited, it is an interesting
quantity in that it captures the rate at which spontaneously generated
information is actively stored by neurons. That is, it quantifies how neurons
harness randomness.

Our contributions to this endeavor are more substantial than the preceding
points. We provided exact formulae for the above quantities for renewal
processes and alternating renewal processes. The new expressions can be
developed further as lower bounds and empirical estimators for a process'
statistical complexity, excess entropy, and bound information rate. This
parallels how the renewal-process entropy-rate formula is a surprisingly
accurate entropy-rate estimator \cite{YGao08}. By deriving explicit
expressions, we were able to analyze time-resolution scaling, showing that the
statistical complexity diverges logarithmically for all but Poisson processes.
So, just like the entropy rate, any calculations of the statistical
complexity---e.g., as in Ref. \cite{Shaliz10a}---should be accompanied by the
time discretization dependence. Notably, the excess entropy and the bound
information rate have no such divergences.

To appreciate more directly what neural information processing behavior these
information measures capture in the continuous-time limit, we studied them as
functions of the ISI coefficient of variation. With an appropriate
renormalization, simulations revealed surprising simplicity: a universal
dependence on the coefficient of variation across several familiar neural
models. The simplicity is worth investigating further since the dynamics and
biophysical mechanisms implicit in the alternative noisy integrate-and-fire
neural models are quite different.  If other generative models of neural spike trains also show similar information universality, then these information measures might prove useful as model selection tools.

Finally, we close with a discussion of a practical issue related to the scaling
analyses---one that is especially important given the increasingly
sophisticated neuronal measurement technologies coming online at a rapid pace
\cite{Aliv12a}. How small should $\Delta t$ be to obtain correct estimates of
neuronal communication? First, as we emphasized, there is no single ``correct''
estimate for an information quantity, rather its resolution scaling is key.
Second, results presented here and in a previous study by others
\cite{Nemenman08a} suggest that extracting information scaling rates and
nondivergent components can require submillisecond time resolution. Third, and
to highlight, the regime of infinitesimal time resolution is exactly the limit
in which computational efforts without analytic foundation will fail or, at a
minimum, be rather inefficient. As such, we hope that the results and methods
developed here will be useful to these future endeavors and guide how new
technologies facilitate scaling analysis.


\acknowledgments

The authors thank C. Hillar and R. G. James for helpful comments. They also
thank the Santa Fe Institute for its hospitality during visits. JPC is an SFI
External Faculty member. This material is based upon work supported by, or in
part by, the U.S. Army Research Laboratory and the U. S. Army Research Office
under contracts W911NF-13-1-0390 and W911NF-13-1-0340. SM was funded by a
National Science Foundation Graduate Student Research Fellowship and the U.C.
Berkeley Chancellor's Fellowship.

\appendix

\section{Alternating renewal process information measures}
\label{app:ARPs}

A discrete-time alternating renewal process draws counts from $F_1(n)$, then
$F_2(n)$, then $F_1(n)$, and so on. We now show that the modality and counts
since last event are causal states when $F_1\neq F_2$ almost everywhere and
when neither $F_1$ nor $F_2$ is eventually $\Delta$-Poisson. We present only a
proof sketch.

Two pasts $\past$ and $\past'$ belong to the same causal state when
$\Prob(\Future|\Past=\past) = \Prob(\Future|\Past=\past')$. We can describe the
future uniquely by a sequence of interevent counts $\mathcal{N}_i$, $i\geq 1$,
and the counts till next event $\mathcal{N}'_0$. Likewise, we could describe
the past as a sequence of interevent counts $\mathcal{N}_i$, $i<0$, and the
counts since last event $\mathcal{N}_0-\mathcal{N}'_0$. Let $\mathcal{M}_i$ be
the modality at time step $i$. So, for instance, $\mathcal{M}_0$ is the present
modality.

First, we claim that one can infer the present modality from a semi-infinite
past almost surely. The probability that the present modality is $1$
having observed the last $2M$ events is:
\begin{align*}
\Prob(\mathcal{M}_0=1 & |\mathcal{N}_{-2M:-1} = n_{-2M:-1}) \\
   & = \prod_{i=-1,\text{odd}}^{2M}F_2(n_{i}) F_1(n_{i-1})
  ~.
\end{align*}
Similarly, the probability that the present modality is $2$ having observed the last $2M$ events is:
\begin{align*}
\Prob(\mathcal{M}_0=2 & |\mathcal{N}_{-2M:-1}=n_{-2M:-1}) \\
   & = \prod_{i=-1,\text{odd}}^{2M}F_1(n_{i}) F_2(n_{i-1})
   ~.
\end{align*}
We are better served by thinking about the normalized difference of the corresponding log likelihoods:
\begin{align*}
Q := \frac{1}{2M} \log
  \frac{P(\mathcal{M}_0=1|\mathcal{N}_{-2M:-1}=n_{-2M:-1})}
  {P(\mathcal{M}_0=2|\mathcal{N}_{-2M:-1}=n_{-2M:-1})}
  ~.
\end{align*}
Some manipulation leads to:
\begin{align*}
Q & = \frac{1}{2}
  \Big( \frac{1}{M}\sum_{i=-1,\text{odd}}^{2M} \log \frac{F_2(n_i)}{F_1(n_i)}
  + \frac{1}{M}\sum_{i=-1,\text{even}}^{2M} \log \frac{F_1(n_i)}{F_2(n_i)}  \Big),
\end{align*}
and, almost surely in the limit of $M\rightarrow\infty$:
\begin{equation}
\frac{1}{M}\sum_{i=-1,\text{odd}}^{2M} \log \frac{F_1(n_i)}{F_2(n_i)} \rightarrow \begin{cases}
  D[F_2||F_1]  & \mathcal{M}_0 = 1 \\
  -D[F_1||F_2] & \mathcal{M}_0 = 2
  \end{cases}
  ~,
\end{equation}
where $D[P || Q]$ is the information gain between $P$ and $Q$
\cite{Cove06a}.
And, we also have:
\begin{align*}
\frac{1}{M}\sum_{i=-1,\text{even}}^{2M} \log \frac{F_2(n_i)}{F_1(n_i)} \rightarrow
  \begin{cases}
  -D[F_1||F_2] & \mathcal{M}_0 = 1 \\
  D[F_2||F_1] & \mathcal{M}_0 = 2
  \end{cases}
  ~.
\end{align*}
This implies that:
\begin{align*}
\lim_{M\rightarrow\infty} Q = \frac{D[F_2||F_1]-D[F_1||F_2]}{2}
  \begin{cases}
  1 & \mathcal{M}_0 = 1 \\
  -1 & \mathcal{M}_0 = 2
  \end{cases}
  ~.
\end{align*}
We only fail to identify the present modality almost surely from the
semi-infinite past if $\lim_{M\rightarrow\infty}Q=0$. Otherwise, the
unnormalized difference of the log likelihoods:
\begin{align*}
\log \frac{\Prob(\mathcal{M}_0=1|\mathcal{N}_{:-1}=n_{:-1})}
          {\Prob(\mathcal{M}_0=2|\mathcal{N}_{:-1}=n_{:-1})}
\end{align*}
tends to $\pm\infty$, implying that one of the two probabilities has
vanished. From the expression, $\lim_{M\rightarrow\infty}Q=0$ only happens when
$D[F_2||F_1] = D[F_1||F_2]$. However, equality requires that $F_1(n)=F_2(n)$ almost everywhere.

Given the present modality, we also need to know the counts since the last
event in order to predict the future as well as possible.  The proof of this is
very similar to those given in Ref. \cite{Marz14b}. The conditional probability
distribution of future given past is:
\begin{align*}
\Prob(\Future|\Past=\past)
  = \Prob(\mathcal{N}_{1:} & |\mathcal{N}_0,\Past=\past) \\
  & \Prob(\mathcal{N}_0|\Past=\past)
  ~.
\end{align*}
Since the present modality is identifiable from the past $\past$, and since interevent counts are independent given modality:
\begin{align*}
\Prob(\mathcal{N}_{1:}|\mathcal{N}_0,\Past=\past)
  = \Prob(\mathcal{N}_{1:}|M_0 = m_0(n_{:-1}))
  ~.
\end{align*}
So, it is necessary to know the modality in order to predict the future as well as possible. By virtue of how the alternating renewal process is generated,
the second term is:
\begin{align*}
\Prob(\mathcal{N}_0|\Past=\past)
  = \Prob(\mathcal{N}_0 | \mathcal{N}_0' = n_0',M_0 = m_0(n_{:-1}))
  ~.
\end{align*}
A very similar term was analyzed in Ref. \cite{Marz14b}, and that analysis revealed that it was necessary to store the counts since last spike when neither $F_1$ nor $F_2$ is eventually $\Delta$-Poisson.

Identifying causal states $\St^+$ as the present modality
$\mathcal{M}_0$ and the counts since last event $\mathcal{N}_0'$ immediately
allows us to calculate the statistical complexity and entropy rate. The
entropy rate can be calculated via:
\begin{align*}
\hmu & = H[\MS_0|\mathcal{M}_0,\mathcal{N}_0'] \\
     & = \pi(\mathcal{M}_0 = 1) H[\MS_0|\mathcal{M}_0=1,\mathcal{N}_0'] \\
     & \quad\quad + \pi(\mathcal{M}_0 = 2) H[\MS_0|\mathcal{M}_0=2,\mathcal{N}_0']
  ~.
\end{align*}
The statistical complexity is:
\begin{align}
\Cmu & = H[\FutureCausalState] \nonumber \\
     & = H[\mathcal{M}_0,\mathcal{N}_0'] \nonumber \\
     & = H[\mathcal{M}_0] + \pi (\mathcal{M}_0 = 1)
	 H[\mathcal{N}_0'|\mathcal{M}_0 = 1] \nonumber \\
     & \quad\quad\quad + \pi (\mathcal{M}_0 = 2) H[\mathcal{N}_0'|\mathcal{M}_0 = 2]
	 ~.
\label{eq:CmuARPviaEntropies}
\end{align}

Finally, it is straightforward to show that the modality $\mathcal{M}_1$ at time step $1$ and the counts to next event are the reverse-time causal states under the same conditions on $F_1$ and $F_2$. Therefore:
\begin{align*}
\EE & = I[\FutureCausalState;\PastCausalState] \\
    & = I[\mathcal{M}_0,\mathcal{N}_0';\mathcal{M}_1,\mathcal{N}_0-\mathcal{N}_0'] \\
    & = I[\mathcal{M}_0;\mathcal{M}_1,\mathcal{N}_0-\mathcal{N}_0'] \\
    & \quad\quad\quad + I[\mathcal{N}_0';\mathcal{M}_1,\mathcal{N}_0-\mathcal{N}_0'|\mathcal{M}_0]
	~.
\end{align*}
One can continue in this way to find formulae for other information measures of a
discrete-time alternating renewal process.

These formulae can be rewritten terms of the modality-dependent information
measures of Eqs.~(\ref{eq:ARPmu})-(\ref{eq:ARPEE}) if we recognize two things.
First, the probability of a particular modality is proportional to the average
amount of time spent in that modality. Second, for reasons similar to those
outlined in Ref. \cite{Marz14b}, the probability of counts since last event
given a particular present modality $i$ is proportional to $w_i(n)$. Hence, in
the infinitesimal time discretization limit, the probability of modality $1$
is:
\begin{align*}
\pi(\mathcal{M}_0 = 1) = \frac{\mu^{(1)}}{\mu^{(1)}+\mu^{(2)}}
\end{align*}
and similarly for modality $2$.
Then, the entropy rate out of modality $i$ is:
\begin{align*}
H[\MS_1|\mathcal{M}_0=i,\mathcal{N}_0'] \sim \Delta t
   \left( \mu^{(i)}\log_2\frac{1}{\Delta t} + \hmu^{(i)}(\Delta t) \right)
  ~,
\end{align*}
and the modality-dependent statistical complexity diverges as:
\begin{align*}
H[\mathcal{N}_0'|\mathcal{M}_0 = i] \sim \log_2 1/\Delta t + \Cmu(\Delta t)
  ~.
\end{align*}
Finally, in continuous-time $\mathcal{M}_0$ and $\mathcal{M}_1$ limit to the
same random variable, such that:
\begin{align*}
\lim_{\Delta t \to 0}\EE(\Delta t) = H[\mathcal{M}_0]
   + \lim_{\Delta t\rightarrow 0}
   I[\mathcal{N}_0';\mathcal{N}_0-\mathcal{N}_0'|\mathcal{M}_0]
   ~.
\end{align*}
Note that $\EE^{(i)}=\lim_{\Delta t\rightarrow 0}
I[\mathcal{N}_0';\mathcal{N}_0-\mathcal{N}_0'|\mathcal{M}_0=i]$.

Bringing these results together, we substitute the above components into Eq.
(\ref{eq:CmuARPviaEntropies})'s expression for $\Cmu$ and, after details not
shown here, find the expression quoted in the main text as Eq.
(\ref{eq:ARPCmu}). Similarly, for $\hmu$ and $\EE$, yielding the the formulae
presented in the main text in Eqs.~(\ref{eq:ARP1}) and (\ref{eq:ARPEnd}),
respectively.

As a last task, as our hypothetical null model, we wish to find the information
measures for the corresponding renewal process approximation.  The ISI
distribution of the alternating renewal process is:
\begin{equation}
\ISI (t) = \frac{ \mu^{(2)} \ISI^{(1)}(t) + \mu^{(1)} \ISI^{(2)}(t)}
  {\mu^{(1)} + \mu^{(2)}}
\end{equation}
and its survival function is:
\begin{equation}
\Phi (t) = \frac{ \mu^{(2)} \Phi^{(1)}(t) + \mu^{(1)} \Phi^{(2)}(t)}
  {\mu^{(1)} + \mu^{(2)}}
  ~.
\end{equation}
Hence, its mean firing rate is:
\begin{equation}
\mu = \frac{ \mu^{(1)} + \mu^{(2)} }
{ \mu^{(2)}/\mu^{(1)} + \mu^{(1)}/\mu^{(2)} }
  ~.
\end{equation}
From Sec. \ref{sec:TimeRes}, the entropy rate of the corresponding renewal
process is:
\begin{align*}
\frac{\hmu^{\text{ren}} (\Delta t)}{\Delta t} \sim \mu\log_2 \frac{1}{\Delta t} + \mu H[\ISI(t)]
  ~;
\end{align*}
compare Eq. (\ref{eq:ARP1}).
And, the statistical complexity of the corresponding renewal process is:
\begin{align*}
\Cmu^{\text{ren}} (\Delta t) \sim \log_2\frac{1}{\Delta t} + H[\mu\Phi(t)]
  ~.
\end{align*}
The rate of divergence of $\Cmu^{\text{ren}}(\Delta t)$ is half the rate of
divergence of the true $\Cmu(\Delta t)$, as given in Eq. (\ref{eq:ARPCmu}).
Trivial manipulations, starting from $0\leq \left(
\frac{1}{\mu^{(1)}}-\frac{1}{\mu^{(2)}} \right)^2$, imply that the rate of
entropy-rate divergence is always less than or equal to the mean firing rate
for an alternating renewal process. Jensen's inequality implies that each of
the nondivergent components of these information measures for the renewal
process is less than or equal to that of the alternating renewal process. The
Data Processing Inequality \cite{Cove06a} also implies that the excess entropy
calculated by assuming a renewal process is a lower bound on the true process'
excess entropy.

\section{Simplicity in Complex Neurons}
\label{app:SiCM}

Recall that our white noise-driven linear leaky integrate-and-fire (LIF) neuron
has governing equation:
\begin{equation}
\dot{V} = b -V + a\eta(t)
 ~,
\label{eq:NLLIFapp}
\end{equation}
and, when $V=1$, a spike is emitted and $V$ is instantaneously reset to $0$. We
computed ISI survival functions from empirical histograms of $10^5$ ISIs.
These ISIs were obtained by simulating Eq.~(\ref{eq:NLLIFapp}) in Python/NumPy
using an Euler integrator with time discretization of $1/1000$ of
$\log b / (b-1)$, which is the ISI in the noiseless limit.

The white noise-driven quadratic integrate-and-fire (QIF) neuron has governing
equation:
\begin{equation}
\dot{V} = b + V^2 + a\eta(t)
  ~,
\label{eq:NQIFapp}
\end{equation}
and, when $V=100$, a spike is emitted and $V$ is instantaneously reset to
$-100$. We computed ISI survival functions also from empirical histograms of
trajectories with $10^5$ ISIs. These ISIs were obtained by simulating
Eq.~(\ref{eq:NQIFapp}) in Python/NumPy using an Euler stochastic integrator
with time discretization of $1/1000$ of $\sqrt{\pi/b}$, which is the ISI in the
noiseless limit when threshold and reset voltages are $+\infty$ and $-\infty$,
respectively.

Figure \ref{fig:Universality} shows estimates of the following continuous-time
information measures from this simulated data as they vary with mean firing
rate $\mu$ and ISI coefficient of variation $C_V$. This required us to estimate
$\mu$, $C_V$, and:
\begin{align*}
\Cmu^{CT} &:= \lim_{\Delta t\rightarrow 0} \Cmu(\Delta t) + \log_2 \Delta t
  ~,\\
\EE^{CT}  &:= \lim_{\Delta t\rightarrow 0} \EE(\Delta t) 
  ~,\\
\hmu^{CT} &:= \lim_{\Delta t\rightarrow 0} \frac{\hmu(\Delta t)}{\Delta t} +\mu\log_2 \Delta t 
  ~, \text{~and}\\
\bmu^{CT} &:= \lim_{\Delta t\rightarrow 0} \frac{\bmu(\Delta t)}{\Delta t}
  ~,
\end{align*}
where the superscript $CT$ is a reminder that these are appropriately
regularized information measures in the continuous-time limit.

We estimated $\mu$ and $C_V$ using the sample mean and sample coefficient of
variation with sufficient samples so that error bars (based on studying errors
as a function of data size) were negligible. The information measures required
new estimators, however. From the formulae in Sec. \ref{sec:TimeRes}, we see that:
\begin{align}
\Cmu^{CT} &= \log_2\frac{1}{\mu} - \mu\int_0^{\infty} \Phi(t)\log_2\Phi(t) dt
  ~,\\
\EE^{CT} &= \int_0^{\infty} \mu t\ISI(t)\log_2(\mu\ISI(t)) dt \nonumber \\
  & \quad\quad - 2\int_0^{\infty} \mu \Phi(t)\log_2\Phi(t) dt
  ~,\\
\hmu^{CT} &= - \mu \int_0^{\infty} \ISI(t)\log_2\ISI(t)
  ~, \text{~and}\\
\bmu^{CT} &= -\mu \Big( \int_0^{\infty} \ISI(t)\int_0^{\infty} \ISI(t') \log_2 \ISI(t+t') dt' dt \nonumber \\
& \quad\quad + \frac{1}{\log 2} - \int_0^{\infty} \ISI(t)\log_2 \ISI(t) dt \Big)
  ~.
\end{align}
It is well known that the sample mean is a consistent estimator of the true
mean, that the empirical cumulative density function is a consistent estimator
of the true cumulative density function almost everywhere, and thus that the
empirical ISI distribution is a consistent estimator of the true cumulative
density function almost everywhere.  In estimating the empirical cumulative
density function, we introduced a cubic spine interpolator. This is still a
consistent estimator as long as $\Phi(t)$ is three-times differentiable, which
is the case for ISI distributions from integrate-and-fire neurons.
We then have estimators of $\Cmu^{CT}$, $\EE^{CT}$,
$\hmu^{CT}$, and $\bmu^{CT}$ that are based on consistent estimators of $\mu$,
$\Phi(t)$, and $\ISI(t)$ and that are likewise consistent.

We now discuss the finding evident in Fig. \ref{fig:Universality}, that the
quantities $\lim_{\Delta t\rightarrow 0}\EE(\Delta t)$ and $\lim_{\Delta
t\rightarrow 0} \Cmu+\log_2 (\mu\Delta t)$ depend only on the ISI coefficient
of variation $C_V$ and not the mean firing rate $\mu$.  Presented in a
different way, this is not so surprising. First, we use Ref.
\cite{Marz14b}'s expression for $\Cmu$
to rewrite:
\begin{align*}
Q_1 & = \lim_{\Delta t\rightarrow 0}
   \left( \Cmu(\Delta t)+\log_2 (\mu\Delta t) \right) \\
    & = - \mu \int_0^{\infty} \Phi(t)\log_2 \Phi(t) dt
\end{align*}
and Eq.~(\ref{eq:EE_CT}) to rewrite:
\begin{align*}
Q_2 & = \lim_{\Delta t\rightarrow 0} \EE(\Delta t) \\
    & = 2Q_1 + \int_0^{\infty} \mu t \ISI(t) \log_2 (\mu \ISI(t)) dt
	~.
\end{align*}
So, we only need to show that $- \mu \int_0^{\infty} \Phi(t)\log_2 \Phi(t) dt$ and $\int_0^{\infty} \mu t \ISI(t) \log_2 (\mu \ISI(t)) dt$ are independent of $\mu$ for two-parameter families of ISI distributions.

Consider a change of variables from $t$ to $t' = \mu t$; then:
\begin{eqnarray}
Q_1 &=& -\int_0^{\infty} \Phi \left( t'/\mu \right)
	\log_2 \big( \Phi \left( t'/\mu \right) \big) dt'
\end{eqnarray}
and
\begin{eqnarray}
Q_2 &=& 2Q_1 + \int_0^{\infty} t' \ISI \left( t'/\mu \right)
   \log_2 \big( \ISI \left( t'/\mu \right) \big) dt'
   ~.
\end{eqnarray}
For all of the ISI distributions considered here, $\ISI \left(\frac{t'}{\mu}
\right)$ is still part of the same two-parameter family as $\ISI(t)$, except
that its mean firing rate is $1$ rather than $\mu$. Its $C_V$ is unchanged. Hence, $Q_1$ and $Q_2$ are the same for a renewal process with mean
firing rate $1$ and $\mu$, as long as the $C_V$ is held constant. It follows
that $\lim_{\Delta t\rightarrow 0}\EE(\Delta t)$ and $\lim_{\Delta t\rightarrow
0} \Cmu+\log_2 (\mu\Delta t)$ are independent of $\mu$ and only depend on
$C_V$ for the two-parameter families of ISI distributions considered in Sec.
\ref{sec:ComplexityMetrics}. Similar arguments apply to understanding the
universal $C_V$-dependence of $\lim_{\Delta t\rightarrow 0} \bmu(\Delta t)/\mu\Delta t$
and $\lim_{\Delta t\rightarrow 0} \hmu(\Delta t)/\mu\Delta t + \log_2 (\mu\Delta t)$.

In Fig. \ref{fig:Universality}, we also see that $\EE$ seems to
diverge as $C_V\rightarrow 0$. Consider the following plausibility argument
that suggests it diverges as $\log_2 1 / C_V$ as $C_V\rightarrow 0$. These
two-parameter ISI distributions with finite mean firing rate $\mu$ and small
$C_V \ll 1$ can be approximated as Gaussians with mean $1/\mu$ and standard
deviation $C_V/\mu$. Recall from Eq.~(\ref{eq:EE_CT}) that we have:
\begin{align*}
\EE & = -2\int_0^{\infty} \mu\Phi(t)\log_2(\mu\Phi(t)) dt \nonumber \\
    & \quad\quad + \int_0^{\infty} \mu t \ISI(t)\log_2 (\mu \ISI(t)) dt \\
    & = -\log_2 \mu - 2\mu \int_0^{\infty} \Phi(t)\log_2 \Phi(t) dt \nonumber \\
    & \quad\quad + \mu \int_0^{\infty} t \ISI(t)\log_2 \ISI(t) dt
  ~.
\end{align*}
Note that as $C_V\rightarrow 0$:
\begin{align}
\Phi(t) \rightarrow
  \begin{cases}
  1 & t<\frac{1}{\mu} \\ \frac{1}{2} & t=\frac{1}{\mu} \\ 0 & t>\frac{1}{\mu}
  \end{cases}
\label{eq:StepFnctn}
\end{align}
and so:
\begin{align*}
\lim_{C_V\rightarrow 0} \int_0^{\infty} \Phi(t)\log_2 \Phi(t) dt = 0
  ~.
\end{align*}
We assumed that for small $C_V$, we can approximate:
\begin{align*}
\ISI(t) \approx \frac{1}{\sqrt{2\pi C_V^2/\mu^2}}
  \exp\left(-\frac{(\mu t-1)^2}{2C_V^2}\right)
  ~,
\end{align*}
which then implies that:
\begin{equation}
\mu \int_0^{\infty} t \ISI(t)\log_2 \ISI(t) dt \approx \log_2 \frac{\mu\sqrt{2\pi}}{C_V} - \frac{1}{2}.
\end{equation}
So, for any ISI distribution tightly distributed about its mean ISI, we expect:
\begin{align*}
\EE \approx  \log_2\frac{1}{C_V}
  ~,
\end{align*}
so that $\EE$ diverges in this way.
A similar asymptotic analysis also shows that as $C_V\rightarrow 0$,
\begin{equation}
\lim_{\Delta t\rightarrow 0}\frac{\bmu(\Delta t)}{\Delta t} \approx \frac{1}{\log 2}(\frac{1}{2C_V^2}-\frac{1}{2})
  ~,
\end{equation}
thereby explaining the divergence of $\lim_{\Delta t\rightarrow 0} \bmu(\Delta
t)/\Delta t$ evident in Fig.~\ref{fig:Universality}.

Finally, a straightforward argument shows that $\Cmu$ is minimized at fixed
$\mu$ when the neural spike train is periodic.  We can rewrite $\Cmu$ in the
infinitesimal time resolution limit as:
\begin{align*}
\Cmu(\Delta t) \sim \log_2 \left(\frac{1}{\mu\Delta t}\right) + \mu \int_0^{\infty} \Phi(t)\log_2 \frac{1}{\Phi(t)}dt
  ~.
\end{align*}
Note that $0\leq \Phi(t)\leq 1$, and
so $\int_0^{\infty} \Phi(t)\log_2 \frac{1}{\Phi(t)}dt\geq 0$. We set it equal
to zero by using the step function given in Eq.~(\ref{eq:StepFnctn}), which
corresponds to a noiseless periodic process. So, the lower bound on
$\Cmu(\Delta t)$ is $\log_2 1 / \mu\Delta t$, and this bound is achieved by a
periodic process.

\bibliography{chaos}

\end{document}